\documentclass[preprint,10.5pt,onside,a4paper]{article}

\usepackage{amssymb}
\usepackage{amsmath}
\usepackage{amscd}
\usepackage{amsthm}
\usepackage{enumerate}
\usepackage[pdftex]{graphicx,color}
\usepackage{lscape}
\usepackage{multirow}
\usepackage{url}
\usepackage{authblk}
\newtheorem{theorem}{Theorem}

\newcommand{\IE}{\textit{i.e., }}
\newcommand{\CF}{\textit{cf.\ }}	% Compare to
\newcommand{\EG}{\textit{e.g., }}

\newcommand{\ETAL}{\textit{et.\ al.\ }}

\newcommand{\abs}[1]{ {\left\lvert #1 \right\rvert} }

\begin{document}                  % DO NOT DELETE THIS LINE

     %-------------------------------------------------------------------------
     % The introductory (header) part of the paper
     %-------------------------------------------------------------------------

     % The title of the paper. Use \shorttitle to indicate an abbreviated title
     % for use in running heads (you will need to uncomment it).

\title{Distribution rules of systematic absence and generalized de Wolff figure of merit applied to EBSD ab-initio indexing}
%\shorttitle{Systematic absence rules and de Wolff $M$ applied to EBSD indexing}

     % Authors' names and addresses. Use \cauthor for the main (contact) author.
     % Use \author for all other authors. Use \aff for authors' affiliations.
     % Use lower-case letters in square brackets to link authors to their
     % affiliations; if there is only one affiliation address, remove the [a].

\author[a,b]{R. Oishi-Tomiyasu}
\author[c]{T. Tanaka}
\author[c]{J. Nakagawa}

\affil[a]{Yamagata University, Yamagata, Japan\thanks{E-mail: tomiyasu@imi.kyushu-u.ac.jp, Current affiliation: Institute of Mathematics for Industry (IMI), Kyushu University}}
\affil[b]{JST PRESTO, Kawaguchi, Japan}
\affil[c]{Nippon Steel Corporation, Japan}

     % Use \shortauthor to indicate an abbreviated author list for use in
     % running heads (you will need to uncomment it).

%\shortauthor{Oishi-Tomiyasu \textit{et. al.}}

     % Use \vita if required to give biographical details (for authors of
     % invited review papers only). Uncomment it.

% \vita{Joe Soape is an archetypal generic author, whose association with the
% much-travelled Kilroy has extended over many years. He travels to work each
% day on a Clapham omnibus.
% \\
% John Doe is also a generic individual with extensive experience of legal and
% forensic matters.}

     % Keywords (required for Journal of Synchrotron Radiation only)
     % Use the \keyword macro for each word or phrase, e.g. 
     % \keyword{X-ray diffraction}\keyword{muscle}

%\keyword{ab-initio indexing}
%\keyword{figure of merit}
%\keyword{electron backscatter diffraction}
%\keyword{EBSD}
%\keyword{Kikuchi pattern}

     % PDB and NDB reference codes for structures referenced in the article and
     % deposited with the Protein Data Bank and Nucleic Acids Database (Acta
     % Crystallographica Section D). Repeat for each separate structuree.g.
     % \PDBref[dethiobiotin synthetase]{1byi} \NDBref[d(G$_4$CGC$_4$)]{ad0002}

%\PDBreference[optional name]{refcode}
%\NDBreference[optional name]{refcode}

\maketitle                        % DO NOT DELETE THIS LINE

%\begin{synopsis}
%A new method for EBSD ab-initio indexing is reported. 
%As a new sorting criterion for the solutions of Kikuchi-pattern indexing, generalized de Wolff figures of merit are also introduced.
%\end{synopsis}

\begin{abstract}
For EBSD ab-initio indexing,  a new method that adopts several methods originally invented for powder indexing, is reported.
Distribution rules of systematic absence and error-stable Bravais lattice determination
are used 
to eliminate negative influence of non-visible bands and erroneous information from visible bands.
In addition, generalized versions of the de Wolff figures of merit are proposed as a new sorting criterion for the obtained unit-cell parameters,
which can be used in both orientation determination and ab-initio indexing from Kikuchi patterns.
Computational results show that the new figures of merit work well, similarly to the original de Wolff $M_n$.
Ambiguity of indexing solutions is also pointed out, which happens in particular for low-symmetric cells, and may generate multiple distinct solutions even if very accurate positions of band center lines and the projection center are given.
\end{abstract}

     %-------------------------------------------------------------------------
     % The main body of the paper
     %-------------------------------------------------------------------------
     % Now enter the text of the document in multiple \section's, \subsection's
     % and \subsubsection's as required.

\section{Introduction}
\label{Introduction}

Electron backscatter diffraction (EBSD) is a characterization technique for the microstructure of crystalline or polycrystalline materials,
developed by Venable \& Harland (1973)\nocite{Venables1973}, and later refined by Dingley \& Baba-Kishi (1986) 
with an aid of computers\nocite{Dingley1986}.
This technique can be applied for the determination of crystal orientation, texture analysis, phase identification
and strain analysis (Troost \ETAL (1993); Wilkinson \ETAL (2006); Tanaka \ETAL (2019)).
\nocite{Troost93}\nocite{Wilkinson2006}\nocite{Tanaka2019}

In orientation determination, the unit-cell parameters are priorly given. 
The center lines of the Kikuchi bands are utilized for acquisition of the unit-cell orientation (Wright \& Adams, 1992; Kogure, 2003\nocite{Wright92}\nocite{Kogure2003}).
However, in EBSD ab-initio indexing, the unit-cell parameters and its symmetry are also determined.
For any fixed $h k \ell$, all the bands with the indices $m (hk\ell)$ ($m$: integer) completely overlap in EBSD patterns.
Hence, all the derivative lattices (\EG sublattices and superlattices) of the true solution have identical positions of Kikuchi center lines.
As a result, 
reliable information about $d$-spacings contained \EG in the widths of Kikuchi bands is indispensable for uniquely determining the solution.
%although the unit-cell parameters and symmetries of these lattices are distinct from those of the true ones.
%It is possible to resolve this ambiguity by choosing the solution with the highest symmetry, only when the cell has cubic symmetry.

The band width is approximately proportional to the inverse of the interplanar spacing (\IE $d$-spacings) of the diffracting plane.
This information has been used for EBSD ab-initio indexing (Michael (2000)\nocite{Michael2000a}; Dingley \& Wright (2009)\nocite{Dingley2009}; Li \& Han (2015)\nocite{Li2015}), 
including the recent software EBSDL \cite{Li2014}.

Due to the complex profile of band edges (Nolze \textit{et. al.}(2015)\nocite{Nolze2015}; Nolze \& Winkelmann (2017)\nocite{Nolze2017a}), and also due to the small Bragg angle caused by short wavelengths of the incident electron beam, the error in the band-width measurement is 5-20\% \cite{Dingley2009}.
%In order to improve the accuracy of measurement of the $d$-spacings, Michael \& Eades (2000) \nocite{Michael2000b} analyzed higher-order Laue Zone (HOLZ) rings observed in EBSD patterns. 
Better accuracy could be obtained by analyzing the higher-order Laue Zone (HOLZ) rings %, compared to the analysis of bandwidths
(Michael \& Eades (2000); Langer \& D{\" a}britz (2007)\nocite{Langer2007};  Dingley \& Wright (2009)\nocite{Dingley2009}).
However, it is not straightforward to analyze the $d$-spacings from the HOLZ rings \cite{Nolze2015}, 
%because pattern simulation based on the dynamical theory is required for reasonable prediction of the HOLZ rings.
and it depends on the crystal structure whether the HOLZ rings are clearly visible.

Therefore, the $d$-spacings extracted from the bandwidths are also used in this article, although
our improvements can be similarly applied to the analyses based on HOLZ rings.

%As a feature of our software, it allows errors in the projection center, band positions and band edges to some degree, as seen from the input file (the first parameter in Table~1 of Appendix~A). 
%In particular, the necessity of a precise projection center has been pointed out \EG in Nolze \& Winkelmann (2017).
%In Section~\ref{}, it is checked how large projection-center shift is allowed by the software.

Technically, the novel points of our indexing method are as follows:
\begin{enumerate}[(1)]
	\item
Non-visible band edges in EBSD patterns, are frequently caused by reflections with relatively small structure factors \cite{Nolze2017a}.%Thus, our main idea concerning (1) is to apply them to the case of EBSD ab-inito indexing. 
We propose a method that works for all the types of systematic absence (SA), as a result of 
the theorems in Oishi-Tomiyasu (2013)\nocite{Tomiyasu2013} that are available without any prior information on the Bravais types and the space groups.
%In order to consider non-visible bands and band edges, due to various reasons including small intensities and systematic absence,

	\item our method for error-stable Bravais lattice determination \cite{Tomiyasu2012}, and new figures of merit with a definition similar to the de Wolff figure of merit\cite{DeWolff68}, are applied to EBSD indexing for the first time.

\end{enumerate}

As for (1), the method of Dingley \& Wright (2009)
needs a reciprocal-lattice basis $l_1^*$, $l_2^*$, $l_3^*$
such that all of $l_1^*$, $-l_i^*$, $l_1^* + l_i^*$ are not extinct for both $i = 2, 3$, 
although such a basis does not exist for some space groups and settings (\EG No.70 $c, d$, No.88 $c, d$).
It is explained in Section~2 how to 
extract information about non-visible bands from visible bands, without being adversely affected by forbidden reflections.
%In our method, the positions and widths of non-visible bands are predicted from the visible ones, in order to construct 3D lattices.
%Our idea is basically same as that proposed in Ito (1949)\nocite{Ito49} for powder indexing, and later developed for \textit{CONOGRAPH} \cite{Tomiyasu2017}.
However, with regard to SAs, Day (2008)\nocite{Day2008} reported that in the experimental pattern of $Si$,
the bandwidth of the forbidden $\{ 222 \}$ was the most visible among all of $\{ h h h \}$.
Therefore, reflection rules might be violated to some degree owing to dynamic scattering of the electron beam.

%For example,
%the length ratio $\abs{ {\mathbf a}_1^* } : \abs{ {\mathbf a}_2^* } : \abs{ {\mathbf a}_3^* }$ is computed from the band positions, by assuming the equation ${\mathbf a}_3^* = p {\mathbf a}_1^* + q {\mathbf a}_2^*$.
%From a theoretical reason explained in Section 2, the $(p, q)$ may be fixed to either of $(1, 1), (1, 2)$ or $(2, 2)$ in order to reduce the computation time.

With regard to the Bravais lattice determination, even very small errors in unit-cell parameters such as rounding errors 
can cause failure in Bravais lattice determination \cite{Kunstleve2004}.
Owing to this, the first author provided a method for error-stable Bravais lattice determination
with rigorous proofs in Oishi-Tomiyasu (2012)\nocite{Tomiyasu2012}, as explained in Section 3.2.

We also propose figures of merit for orientation determination and ab-initio indexing (Section~4),
by extending the definition of the de Wolff figure of merit.
The de Wolff $M_n$ has been used as the most efficient indicator in powder indexing (1D data),
and the generalized ones are presenting very similar properties to those of the original one.

In this article, an ambiguity of solutions in EBSD ab-initio indexing,
which sometimes allows multiple distinct solutions, are also explained.
This is different from the ambiguity reported in Alkorta (2013)\nocite{Alkorta2013} (explained in Sections 2 and 3.2),
and happens when many observed bandwidths are not the narrowest ones.
In this case, the above uniqueness problem occurs again, 
because all the sublattices of the true crystal lattice can have identical bandwidths, in addition to the band positions. 
Thus, in ab-initio indexing, it is necessary to assume that the edges of the narrowest bands are the most visible at least for several bands. Otherwise, the assumption that the unit cell has higher-symmetric Bravais type can be used to resolve this ambiguity.

Lastly, the developed program and the source codes are available from the web site: \url{http://ebsd-conograph.osdn.jp/InstructionsEBSDConograph.html}.

For definition, the 3-dimensional (3D) lattice $M$ is a \textit{derivative lattice} of another 3D lattice $L$,
if they have a common 3D sublattice $M \cap L$.

\section{Background and formulas for EBSD indexing}

%In this section, the background and relation formulas used for EBSD indexing are introduced for the contents in Section~3.
A general method to gain the unit-cell length-ratios and angles from the center-line positions of the Kikuchi bands 
(in particular without bandwidths) is explained herein.
The used notation is basically the same as that in Kogure (2003)\nocite{Kogure2003}.

\begin{figure}[htbp]
\begin{center}
\scalebox{0.36}{\includegraphics{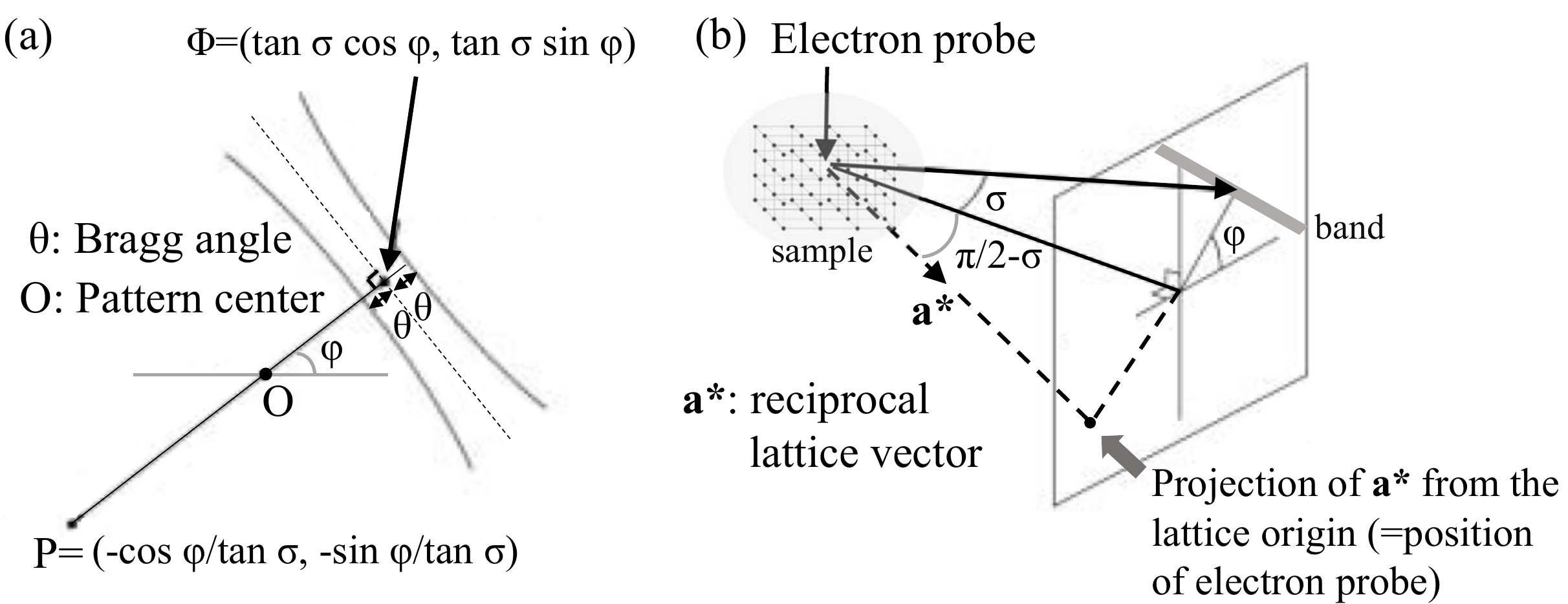}}
\end{center}
\caption{Relationship between the center line of a Kikuchi band and the direction of the corresponding reciprocal lattice vector.
In (a), the phosphor screen is parallel to the sheet. In both (a) and (b),
the scale is adjusted so that the camera length ($=$ distance from the electron probe to the screen) equals 1.
%(a) 
%If the center line and its perpendicular line from the pattern center intersect at   
%$(\tan \sigma, \varphi)$, 
%the line with the direction of the corresponding reciprocal vector ${\mathbf a}^*$ passes through $(1/\tan \sigma, \varphi)$.
%(b) the positions of Kikuchi lines provides the set of the projected points of the reciprocal lattice.
}
\label{fig: transformation_of_EBSD_pattern}
\end{figure}

\begin{figure}[htbp]
\begin{center}
\scalebox{0.45}{\includegraphics{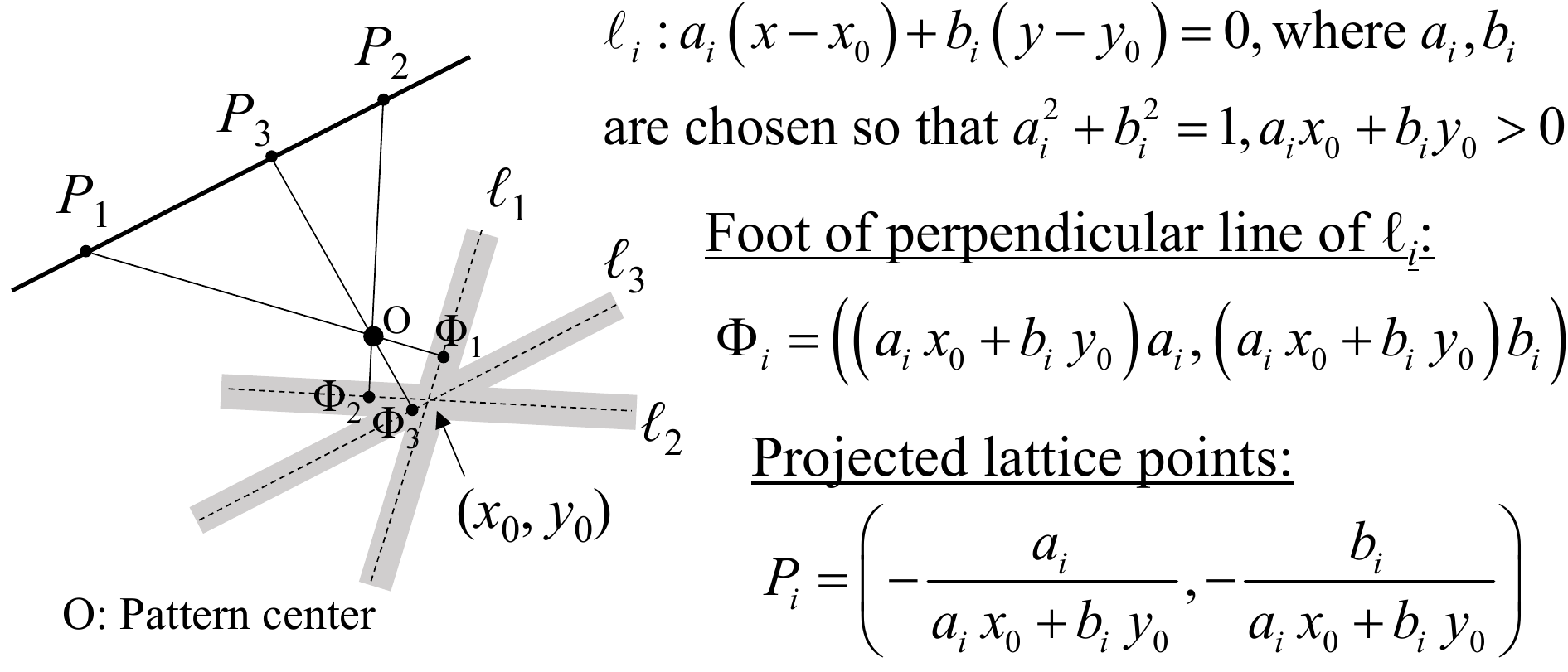}}
\end{center}
\caption{Kikuchi bands intersecting at one point; if all the Kikuchi lines $\ell_i$ ($i=1, 2, 3$) intersect at $(x_0, y_0)$, their corresponding projected lattice points $P_i$ are on the line $x_0 X + y_0 Y = -1$.
}
\label{fig: coplanar_lattice_vector}
\end{figure}

The general situation of electron backscattering is depicted in Figure~\ref{fig: transformation_of_EBSD_pattern}.
In Figures~\ref{fig: transformation_of_EBSD_pattern} and \ref{fig: coplanar_lattice_vector}, 
it may be thought that the positions of the projection centers are exact, even if they are unknown.
Although the relationship between Kikuchi bands is often explained by using the coordinates of Kikuchi bands,
the same thing is more easily understood by using the projected coordinates of the reciprocal lattice points, which are computable from the parameters of the band center lines.

In Figure~\ref{fig: transformation_of_EBSD_pattern} (a), 
the three axes $x_1, x_2, x_3$ orthogonal to each other, are fixed so that
the third axis $x_3$ is perpendicular to the phosphor screen. 
Since the scale is adjusted so that the camera length equals~1, the pattern center $O$ on the phosphor screen exists at the coordinate $(0, 0, 1)$.

The Kikuchi center lines are the intersections of diffracting plains and the phosphor screen.
As in Figure~\ref{fig: transformation_of_EBSD_pattern} (a),
the coordinate of each center lines (more precisely, the foot of the perpendicular from the pattern center $O$ to the Kikuchi center line) can be represented as $(x_1, x_2, x_3) = (\tan \sigma \cos \varphi, \tan \sigma \sin \varphi, 1)$, by using its spherical coordinate $(r, \sigma, \varphi)$ with $r=1/\cos$.
The perpendicular direction  
coincides with the direction of the reciprocal-lattice vectors $m {\mathbf a}^*$ ($m \ne 0$: integer). Namely, 
\begin{eqnarray*}\label{eq: direction of a^*}
{\mathbf a}^* 
	& \propto & \left( - \frac{\cos \varphi}{\tan \sigma}, -\frac{\sin \varphi}{\tan \sigma}, 1 \right), \\
	& \propto & (-\cos \sigma \cos \varphi, -\cos \sigma \sin \varphi, \sin \sigma).
\end{eqnarray*}

As a result, as shown in Figure~\ref{fig: transformation_of_EBSD_pattern}(b),
the reciprocal-lattice vector ${\mathbf a}^*$ that provides the Miller index of the Kikuchi band at $(x_1, x_2) = (\tan \sigma \cos \varphi, \tan \sigma \sin \varphi)$
is projected to the coordinate $(x_1, x_2) = \left(-\cos \varphi/\tan \sigma, -\sin \varphi/\tan \sigma \right)$ on the screen.

This interpretation is useful for obtaining a geometric intuition of Kikuchi patterns.
In particular, 
the bands with the Miller indices $m (hk\ell)$ (equivalently, reciprocal lattice vectors $m {\mathbf a}^*$, $m \ne 0$: integer) have an identical center line. Furthermore, the three projected lattice points $P_1, P_2, P_3$ are aligned on the phosphor screen, if and only if they correspond to coplanar reciprocal lattice vectors.
As is well known, this happens if and only if 
the corresponding Kikuchi lines intersect at one point (Figure~\ref{fig: coplanar_lattice_vector}).

This property can be used when one wants to improve results of automatic band detection.
Under the projection-center shift $(\Delta x, \Delta y, \Delta z)$,
the intersection is varied from $(x_0, y_0)$ to $(x_0 (1-\Delta z) + \Delta x, y_0 (1-\Delta z)+ \Delta y)$
(see Eq.(\ref{eq: definiton of X^{cal}}), (\ref{eq: definiton of Y^{cal}}) to obtain this).

The ratios of the lattice-vector lengths can be determined from the positions of the Kikuchi lines,
as follows;
%For example, in the method of \textit{ITO} (1949)\nocite{Ito49}, the following equation is used to specify the reflections of ${\mathbf a}_1^*$, ${\mathbf a}_2^*$ and ${\mathbf a}_1^* \pm {\mathbf a}_2^*$:
%$$
%2 (\abs{ {\mathbf a}_1^* }^2 + \abs{ {\mathbf a}_2^* }^2 )
%= \abs{ {\mathbf a}_1^* + {\mathbf a}_2^* }^2 + \abs{ {\mathbf a}_1^* - {\mathbf a}_2^* }^2.
%$$
the parameter $[a_i : b_i]$, $a_i^2 + b_i^2 = 1$ taken as in Figure~\ref{fig: coplanar_lattice_vector}, represents the slope of a band.
In this case, the projection $P_i$ of the corresponding reciprocal lattice vector ${\mathbf a}_i^*$ has the coordinate $P_i = (-a_i / (a_i x_0 + b_i y_0), -b_i / (a_i x_0 + b_i y_0))$. 
In the 3D-coordinate system,
the coordinate of $P_i$ is equal to $(-a_i / (a_i x_0 + b_i y_0), -b_i / (a_i x_0 + b_i y_0), 1)$.
Since ${\mathbf a}^*_i$ and $P_i$ have the same direction, for some $c_i > 0$,
$$
	{\mathbf a}^*_i = c_i (-a_i, -b_i, a_i x_0 + b_i y_0).
$$

If ${\mathbf a}_1^*, {\mathbf a}_2^*, {\mathbf a}_3^*$ are coplanar as in Figure~\ref{fig: coplanar_lattice_vector},  
some rational numbers $p, q$ satisfy ${\mathbf a}_3^* = p {\mathbf a}_1^* + q {\mathbf a}_2^*$.
Hence, the ratio $c_1 :c_2 : c_3$ satisfies:
\begin{eqnarray}\label{eq: linear equation to solve}
\begin{pmatrix}
	-a_1 & -a_2 & -a_3 \\
	-b_1 & -b_2 & -b_3 \\
	a_1 x_0 + b_1 y_0 & a_2 x_0 + b_2 y_0 & a_3 x_0 + b_3 y_0
\end{pmatrix}
\begin{pmatrix}
	p c_1\\
	q c_2 \\
	-c_3
\end{pmatrix}
 = 0.
\end{eqnarray}

If the values of $p, q$ are given,
the ratio $c_1 : c_2 : c_3$
can be computed from the inner products $\alpha_{ij} = (a_i, b_i) \cdot (a_j, b_j)$, owing to $a_i^2 + b_i^2 = 1$:
\begin{eqnarray*}
c_1 : c_2 : c_3
&=& \frac{1}{p}
\begin{vmatrix}
	a_2 & a_3 \\
	b_2 & b_3 \\
\end{vmatrix}
: \frac{1}{q}
\begin{vmatrix}
	a_3 & a_1 \\
	b_3 & b_1 \\
\end{vmatrix}
: 
\begin{vmatrix}
	a_1 & a_2 \\
	b_1 & b_2 \\
\end{vmatrix} \\
&=& \frac{\sqrt{1 - \alpha_{23}^2}}{\abs{p}}
: \frac{\sqrt{1 - \alpha_{13}^2}}{\abs{q}}
: \sqrt{1 - \alpha_{12}^2}
\end{eqnarray*}

The slope $[a_i : b_i]$
of a Kikuchi line can be determined independently from the position of the pattern center.
This explains why the obtained $c_1 : c_2 : c_3$ are not affected by the error of the pattern center as well. 
Only the third entry of $a_i x_0 + b_i y_0$ is affected by the shift $(\Delta x, \Delta y)$ of $(x_0, y_0)$.

In EBSD indexing, if $(\Delta x, \Delta y)$ are well refined, even if $\Delta z$ is imprecise,
it is possible to index the band center lines,
although the $z$-scale of the obtained unit-cell parameters might contain large errors owing to the shift $\Delta z$.
This ambiguity was also pointed out in Alkorta (2013) in a special setting.
Eqs.(\ref{eq: definiton of X^{cal}}), (\ref{eq: definiton of Y^{cal}}) in Section~3.3, 
explains how $\Delta z$ and the $z$-scale are corelated, when they are simultaneously determined from an EBSD pattern. 

In ab-initio indexing, although $p$ and $q$ in Eq.(\ref{eq: linear equation to solve}) are unknown, 
the ratio $c_1 : c_2 : c_3$ can be computed by setting $(p, q)$ to specific values \EG $(1, 1)$, (2, 1), or (1, 2) as in Section~\ref{Method for ab-initio EBSD indexing}.

The Kikuchi bandwidth $\beta$ on the screen is related to the Bragg angle $\theta$, as follows:
\begin{eqnarray}\label{eq:band width}
\beta = \tan (\sigma + \theta) - \tan (\sigma - \theta),
\end{eqnarray}
The information about the $d$-spacing ($=1/\abs{ m{\mathbf a}^*}$) of $m {\mathbf a}^*$ and its inverse $d^*$ can be obtained from this $\theta$ by using the Bragg equation: 
\begin{eqnarray}\label{eq:d-values from band width}
	d^* = \frac{1}{d} = \frac{2 \sin \theta}{\lambda},
\end{eqnarray}
where $\lambda$ is the wavelength of the electron beam.
%In what follows, Eq.(\ref{eq:d-values from band width}) is used if the projection-center shift is ignorable.
%In particular, Eq.(\ref{eq:band width}) is used in the refinement process (Section 3.2)

In practice, it is difficult to obtain accurate values of the bandwidths from experimental patterns, and to judge whether the obtained value corresponds to the narrowest band,
although the same thing happens even if HOLZ rings are used \cite{Michael2000b}.
%It may be thought that reflections with relatively large structure factors frequently correspond to visible band widths.

The following theorems 
used in Section~\ref{Method for ab-initio EBSD indexing}, 
state that it is not necessary to assign various sets of reciprocal lattice vectors to a combination of the narrowest bands in order to remove all the adverse effect of SA (\EG, $(p, q) = (2, 1), (1, 2)$ is sufficient  
to obtain various 2D sublattices, if a number of band edges are available).

In the statements of the theorems, $L^*$ is the reciprocal lattice of the crystal lattice $L$.
For simplicity, $L$ always means the primitive lattice (\IE the lattice before centering).
$\{ l_1^*, l_2^* \}$ is called a \textit{ primitive set}, 
if it is a subset of some basis $l_1^*, l_2^*, l_3^*$ of $L^*$.
\begin{theorem}[Theorem 2 in Oishi-Tomiyasu (2013)\nocite{Tomiyasu2013}]\label{thm:construction of 3-dim solution}\label{thm:three-dimensional extinction rules}
Regardless of the type of SA,
there are infinitely many primitive sets $\{ l_1^*, l_2^* \}$ of $L^*$
such that none of $l_1^*$, $l_2^*$, $l_1^* + 2 l_2^*$, $2 l_1^* + l_2^*$ 
corresponds to an extinct reflection due to the SA.
Furthermore, there exist infinitely many 2D sublattices $L_2^*$ of $L^*$ 
such that $L_2^*$ is expanded by such $l_1^*, l_2^*$.
\end{theorem}

The reciprocal lattices $l_1^*$, $l_2^*$, $l_1^* + 2 l_2^*$, $2 l_1^* + l_2^*$, and $l_1^* + l_2^*$ are coplanar.
In the method of \textit{CONOGRAPH}, their relationship is illustrated as in Figure~\ref{A subgraph of a topograph corresponding to the formula 3 abs{ l_1^* }^2 + abs{ l_1^* + 2 l_2^* }^2 = abs{ 2 l_1^* + l_2^* }^2 + 3 abs{ l_2^* }^2}, by using a graph:
\begin{figure}[htbp]
\begin{center}
\scalebox{0.75}{\includegraphics{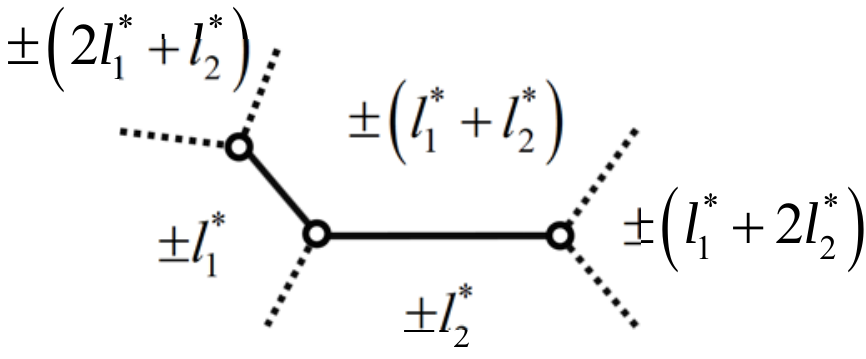}}
\end{center}
\caption{A subgraph of a topograph corresponding to the reflections 
$l_1^*$, $l_2^*$, $l_1^* + 2 l_2^*$, $2 l_1^* + l_2^*$ that are not forbidden, and $l_1^* + l_2^*$ that might be forbidden owing to SA.
This graph was originally used in Conway (1997), where the term ``topograph'' was first coined.
}
\label{A subgraph of a topograph corresponding to the formula 3 abs{ l_1^* }^2 + abs{ l_1^* + 2 l_2^* }^2 = abs{ 2 l_1^* + l_2^* }^2 + 3 abs{ l_2^* }^2}
\end{figure}

\begin{theorem}[Theorem 4 in Oishi-Tomiyasu (2013)\nocite{Tomiyasu2013}]\label{thm:construction of 3-dim solution}
Regardless of the type of SA, there are infinitely many bases $\langle l_1^*, l_2^*, l_3^* \rangle$ of $L^*$ such that the following hold:
\begin{enumerate}[(a)]
		\item the reflections of $\pm l_1^* + l_2^* + l_3^*$ are not forbidden.
		\item \label{item:long subgraph}
			For both $i= 2, 3$, (i) none of the reflections of $m l_1^* + (m-1)(-l_1^* + l_i^*)$ are forbidden for any integer $m$, or otherwise, (ii) none of the reflections of $m l_i^* + (m-1)(l_1^* - l_i^*)$ are forbidden for any integer $m \geq 0$.
\end{enumerate}
\end{theorem}

\begin{figure}[htbp]
\begin{center}
\scalebox{0.37}{\includegraphics{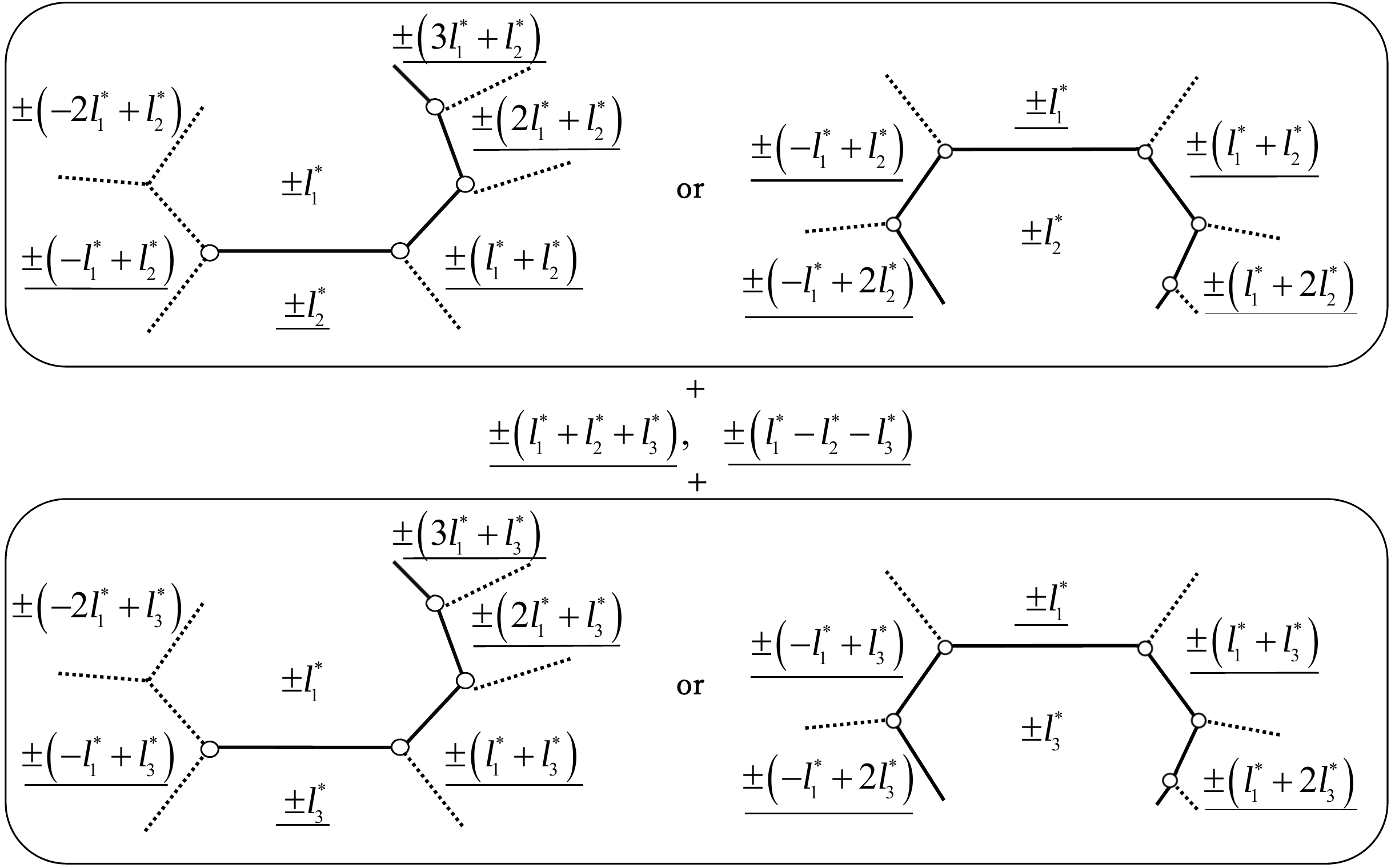}}
\end{center}
\caption{Outline of Theorem \ref{thm:construction of 3-dim solution}, which ensures that none of the underlined lattice vectors are forbidden.}
\label{Materials used to generate 3-dimensional solutions}
\end{figure}

That is, none of the underlined lattice vectors in Figure~\ref{Materials used to generate 3-dimensional solutions} corresponds to a forbidden reflection.

\section{New methods for ab-initio EBSD indexing and scale determination of the unit-cell}
\label{Method for ab-initio EBSD indexing}

\subsection{Acquisition of candidates for the primitive lattice}
\label{Method for ab-initio EBSD indexing (aquisition of candidates for the primitive lattice)}

Ab-initio indexing methods can be classified into two categories, depending on their strategy. In the first category, various $hk\ell$s are assigned to a few selected reflections, in order to generate multiple candidate solutions.
%In the second category, various combinations of reflections are assumed to satisfy some fixed relationship. 
%In both categories, it is checked whether there is a candidate that can well predict all the observed reflections. 
The advantage of the latter is that 
the true solution is normally generated multiple times from distinct observed reflections,
hence it can be robust against errors in the input.

Our method belongs to the latter.
%In what follows, it is also explained how the computation time can be reduced by using the relationship between the observed reflections.
The basic algorithm, which uses only the positions of the Kikuchi center lines,
is provided in Table~\ref{The algorithm for ab-initio EBSD indexing (without information about $d$-spacings}.
%It outputs only a finite number of candidates, thanks to the finiteness of the observed Kikuchi bands,
%although case, all the derivative lattices of the true solution are as good as the true one as indexing solutions.
%In the following paragraphs, methods to reduce the computation time (without \& by using the band widths) are explained.

\begin{table}[htbp]
\caption{Indexing algorithm in which only the positions of Kikuchi center lines are used.
}
\label{The algorithm for ab-initio EBSD indexing (without information about $d$-spacings}
%\begin{tabular}{lp{80mm}}
\begin{minipage}{\textwidth}
\begin{tabular}{lp{130mm}}
\hline
\multicolumn{2}{l}{(Input)} \\
$\mathbf{Inp}$: & array of unit vectors ${\mathbf u} = (-\cos \sigma \cos \varphi, -\cos \sigma \sin \varphi, \sin \sigma)$
obtained from the Kikuchi band center positions $(\tan \sigma \cos \varphi, \tan \sigma \sin \varphi)$ on the screen (the positions may be affected by the error of the projection center). \\

\multicolumn{2}{l}{(Output)} \\
$\mathbf{Ans}$: & array of candidates for the reciprocal lattice basis (here, the basis vectors are the edges of the primitive cell). \\

\multicolumn{2}{l}{(Algorithm)} \\
\multicolumn{2}{l}{{\bf Detection of zones (2D sublattices)}:} \\
(1) & for any distinct vectors ${\mathbf u_1} \neq {\mathbf u_2}$ in $\mathbf{Inp}$,
search for all the ${\mathbf u_3}$ in $\mathbf{Inp}$ 
that may be considered to be linearly dependent on ${\mathbf u_1}, {\mathbf u_2}$.
All of such ${\mathbf u_3}$ are saved in a new array $\mathbf{Inp}_{{\mathbf u_1}, {\mathbf u_2}}$.
%For this part, it is necessary to consider observation errors including those in the pattern center. 
\\
(2) & $\langle$Computation of $\lambda_1$, $\lambda_2$ with ${\mathbf u_3} = \lambda_1 {\mathbf u_1} + \lambda_2 {\mathbf u_2} \rangle$
for each ${\mathbf u_3} \in \mathbf{Inp}_{{\mathbf u_1}, {\mathbf u_2}}$, 
the following equation is solved:
$$
\begin{pmatrix}
	u_{11} & u_{21} & u_{31} \\
	u_{12} & u_{22} & u_{32} 
\end{pmatrix}
\begin{pmatrix}
	\lambda_1 \\ \lambda_2 \\ -1
\end{pmatrix} = 0,
$$
where $(u_{i1}, u_{i2}, u_{i3})$ are the entries of ${\mathbf u}_i$ ($i = 1,2,3$). 
If $\lambda_1 \le 0$ or $\lambda_2 \le 0$, go to the next ${\mathbf u_3} \in \mathbf{Inp}_{{\mathbf u_1}, {\mathbf u_2}}$. 
Otherwise, carry out step (3) and store a pair of vectors $\{ {\mathbf a}_i^*, {\mathbf a}_1^* + {\mathbf a}_2^* \}$ ($i = 1, 2$)
in a common array ${\mathcal A}$, before going to the next ${\mathbf u_3}$. \\
\\

(3) & 
In what follows, ${\mathbf a}_i^*$ is the reciprocal lattice vector with the direction ${\mathbf u}_i$ ($i = 1, 2, 3$).
The assumption ${\mathbf a}_3^* = p {\mathbf a}_1^* + q {\mathbf a}_2^*$ is tested 
for each of $(p, q) = (1, 1), (2, 1), (1, 2)$ in the following (3-a)--(3-c):
\\

(3-a) & (Case of $(p, q) = (1, 1)$, \IE ${\mathbf a}_3^* = {\mathbf a}_1^* + {\mathbf a}_2^*$) in this case, $\abs{ {\mathbf a}_1^* } : \abs{ {\mathbf a}_2^* } : \abs{ {\mathbf a}_3^* } = \lambda_1 : \lambda_2 : 1$ holds (\CF Eq.(\ref{eq: linear equation to solve})). Hence, $\{ \lambda_1 {\mathbf u}_1, {\mathbf u}_3 \}$, $\{ \lambda_2 {\mathbf u}_2, {\mathbf u}_3 \}$,
 are stored in ${\mathcal A}$.
\\ 

(3-b) & (Case of $(p, q) = (2, 1)$, \IE ${\mathbf a}_3^* = 2{\mathbf a}_1^* + {\mathbf a}_2^*$) 
Similarly, $\abs{ {\mathbf a}_1^* } : \abs{ {\mathbf a}_2^* } : \abs{ {\mathbf a}_3^* } = \lambda_1/2 : \lambda_2 : 1$ is obtained. Hence ${\mathbf a}_1^*$, ${\mathbf a}_2^*$, ${\mathbf a}_1^* + {\mathbf a}_2^*$ are constant multiples of $(\lambda_1/2) {\mathbf u}_1$, $\lambda_2 {\mathbf u}_2$, $(\lambda_1/2) {\mathbf u}_1 + \lambda_2 {\mathbf u}_2$.
If the direction of ${\mathbf a}_1^* + {\mathbf a}_2^*$ is not observed (\IE not in $\mathbf{Inp}_{{\mathbf u_1}, {\mathbf u_2}}$), $\{ (\lambda_1/2) {\mathbf u}_1, (\lambda_1/2) {\mathbf u}_1 + \lambda_2 {\mathbf u}_2 \}$, $\{ \lambda_2 {\mathbf u}_2, (\lambda_1/2) {\mathbf u}_1 + \lambda_2 {\mathbf u}_2 \}$ are stored in ${\mathcal A}$.
\\ 

(3-c) & (Case of $\langle (p, q) = (1, 2)$, \IE ${\mathbf a}_3^* = {\mathbf a}_1^* + 2{\mathbf a}_2^*$) in this case, $\abs{ {\mathbf a}_1^* } : \abs{ {\mathbf a}_2^* } : \abs{ {\mathbf a}_3^* } = \lambda_1 : \lambda_2/2 : 1$. Hence ${\mathbf a}_1^*$, ${\mathbf a}_2^*$, ${\mathbf a}_1^* + {\mathbf a}_2^*$ are proportional to $\lambda_1 {\mathbf u}_1$, $(\lambda_2/2) {\mathbf u}_2$, $\lambda_1 {\mathbf u}_1 + (\lambda_2/2) {\mathbf u}_2$.
If the direction of ${\mathbf a}_1^* + {\mathbf a}_2^*$ is not in $\mathbf{Inp}_{{\mathbf u_1}, {\mathbf u_2}}$, $\{ \lambda_1 {\mathbf u}_1, \lambda_1 {\mathbf u}_1 + (\lambda_2/2) {\mathbf u}_2 \}$, $\{ (\lambda_2/2) {\mathbf u}_2, \lambda_1 {\mathbf u}_1 + (\lambda_2/2) {\mathbf u}_2 \}$ are stored in ${\mathcal A}$.
\\ \\

\multicolumn{2}{l}{{\bf Construction of candidates for the lattice basis}:} \\
(4) & for any $\{ {\mathbf b}_1^*, {\mathbf b}_2^* \}$, $\{ c {\mathbf b}_1^*, c{\mathbf b}_3^* \} \in {\mathcal A}$ including vectors ${\mathbf b}_1^*, c{\mathbf b}_1^*$ with the same direction,
	if ${\mathbf b}_1^*$, ${\mathbf b}_2^*$, ${\mathbf b}_3^*$ are linearly independent and pass the following check (*)$^\dagger$, 
	the basis $\{ {\mathbf b}_1^*, {\mathbf b}_2^*, {\mathbf b}_3^* \}$ is stored in $\mathbf{Ans}$ as a candidate solution. \\

& \hspace{7mm}(*)$^{\dagger}$ the direction of ${\mathbf b}_1^* + {\mathbf b}_2^* + {\mathbf b}_3^*$
       is observed, \IE in $\mathbf{Inp}$.\\
\end{tabular}
	\footnotetext[2]{
The check (*) is imposed to reduce the number of solutions and computation time.
By removing (*), it is possible to carry out a more exhaustive search.
	}
\end{minipage}
\end{table}

\begin{figure}[htbp]
\begin{center}
\scalebox{0.47}{\includegraphics{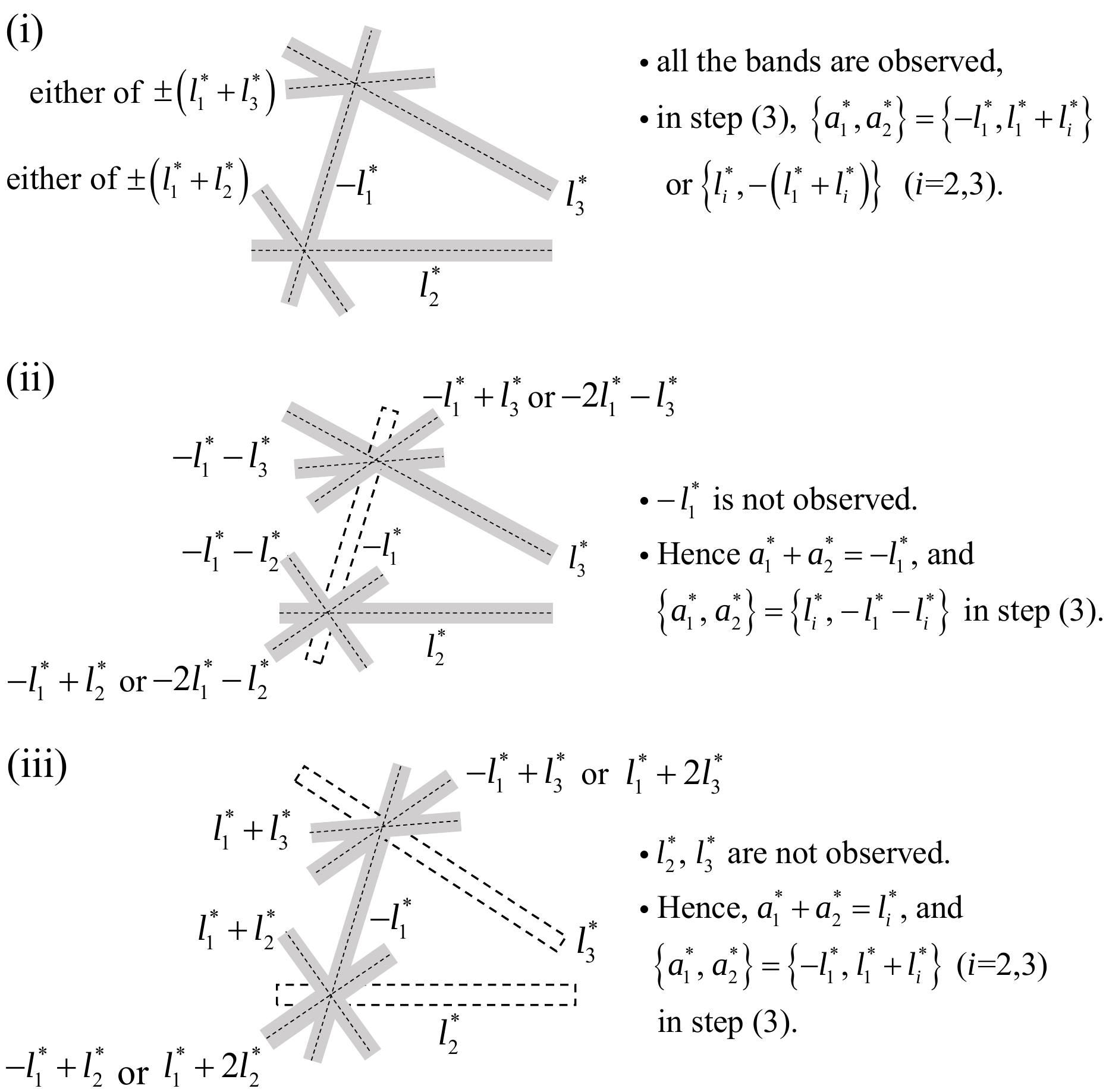}}
\end{center}
\caption{Combinations of bands and their reciprocal lattice vectors assigned for indexing. 
If 5 or 6 bands intersect as in either of (i)--(iii), the lattice basis $({\mathbf b}_1^*, {\mathbf b}_2^*, {\mathbf b}_3^{*}) = (-l_1^*, l_2^*, l_3^* )$ is saved in step (4) of Table~\ref{The algorithm for ab-initio EBSD indexing (without information about $d$-spacings}.
All of their length-ratios and inner products are determined from the band positions.
Every observed band (gray) and computed band (white) are assigned either of $\pm l_1^*$, $\pm l_2^*$, $\pm l_3^*$, or their linear sums $\pm (l_1^* \pm l_i^*)$, $\pm (2 l_1^* + l_i^*)$, $\pm (l_1^* + 2 l_i^*)$ ($i = 2, 3$).
These vectors assigned to the observed bands are chosen from the underlined not-forbidden reflections in Figure~\ref{Materials used to generate 3-dimensional solutions}.
}
\label{fig: combination_of_bands}
\end{figure}

Unlike step (3-a) in which the direction of ${\mathbf a}^*_1 + {\mathbf a}^*_2$ is observed, in steps (3-b) and (3-c), 
${\mathbf a}^*_1 + {\mathbf a}^*_2$ predicted from the other input bands
is stored in ${\mathcal A}$. 
The algorithm is simplified by this use of virtual bands.
%and one of the characteristics of the methods in Table \ref{The algorithm for ab-initio EBSD indexing (without information about $d$-spacings} and in \textit{CONOGRAPH}.
Figure~\ref{fig: combination_of_bands} shows which combinations of visible bands and non-visible bands are used to construct a unit cell. 

So far, although Theorems 1, 2 have been used to determine which sets of vectors should be assigned to bands,  
it is unnecessary to consider SAs,  
because of the overlay of the bands of $m (h k \ell)$ ($m$: integer).
SAs (and the breakdown of Friedel's law \cite{Marthinsen88}) do not largely influence on the band positions. 
%However, it is normally difficult to detect multiple band edges for one trace, especially when experimental patterns are used.
%If at least one of such a lattice basis is found, the algorithm succeeds in ab-initio indexing. 
However, information about bandwidths are used in step (3)
in order to obtain the lengths of ${\mathbf a}_i^*, {\mathbf a}_1^* + {\mathbf a}_2^*$.
SAs influence in this case, because it is assumed that the edges of the narrowest bands are input for the visible bands.
%Subsequently, in step (4), 
%the scale of $\{ {\mathbf b}_1^*, {\mathbf b}_2^*, {\mathbf b}_3^* \}$
%can be calculated from the scales $s_1$, $s_2$ of $\{ {\mathbf b}_1^*, {\mathbf b}_2^* \}$, $\{ c {\mathbf b}_1^*, c{\mathbf b}_3^* \}$,
%for example, as the mean value of $s_1$, $s_2$.
%When each band is indiced to a Miller index $h k \ell$, 
%In the next refinement process (Section~\ref{refinement of pattern center, orientation and unit-cell parameters}),

%If use of the band width is not selected, 
%these additional constraints are not used in our program.

At runtime, our algorithm normally generates multiple times almost identical lattices from different combinations of observed bands.
The unit-cell scales are also computed for each combination.
%Once the unit-cell parameters are determined in this way, 
%it is possible to judge whether or not the other band widths are multiple of the narrowest ones.

\subsection{Bravais lattice determination \& refinement of the projection center, unit-cell parameters and orientation}
\label{refinement of pattern center, orientation and unit-cell parameters}

In Bravais lattice determination, the Bravais-type and the parameters of the conventional cell are determined from the parameters of the primitive cell.
%
%As seen from the notation of the lattice characters in the International Tables Vol.A, 
%this determination can be regarded
%as the process to find lattice vectors that intersect at a specific angle such as $90^\circ$ or $120^\circ$.
%Even if the unit-cell parameters are erroneous, this process can be error-stable,
%although it is necessary to also output parameters relaxed to lower-symmetry cells
%(\EG triclinic and monoclinic cells, if an orthorombic cell is correct). 
%The main problem for software developers is how to find a reliable program.
The method has been studied by using the lattice-basis reduction theory.
For exact unit-cell parameters, the Bravais lattice can be determined by
using the 44 lattice characters (Table 9.2.5.1 of International Tables Vol. A \cite{Hahn2005}).

Oishi-Tomiyasu (2012)\nocite{Tomiyasu2012} reported 
a method to extend the Bravais-lattice determination algorithm to general erroneous cases, without increasing calculation time.
It is guaranteed to output the correct Bravais type and lattice basis of the conventional cell, except for the case when the input cells contain very huge errors (the precise condition was provided in the cited paper).
The program has been used both for determinations under observation errors (\EG \textit{CONOGRAPH} for powder indexing) 
and rounding-off errors\cite{Tomiyasu2016}.

\begin{figure}[htbp]
\begin{center}
\scalebox{0.47}{\includegraphics{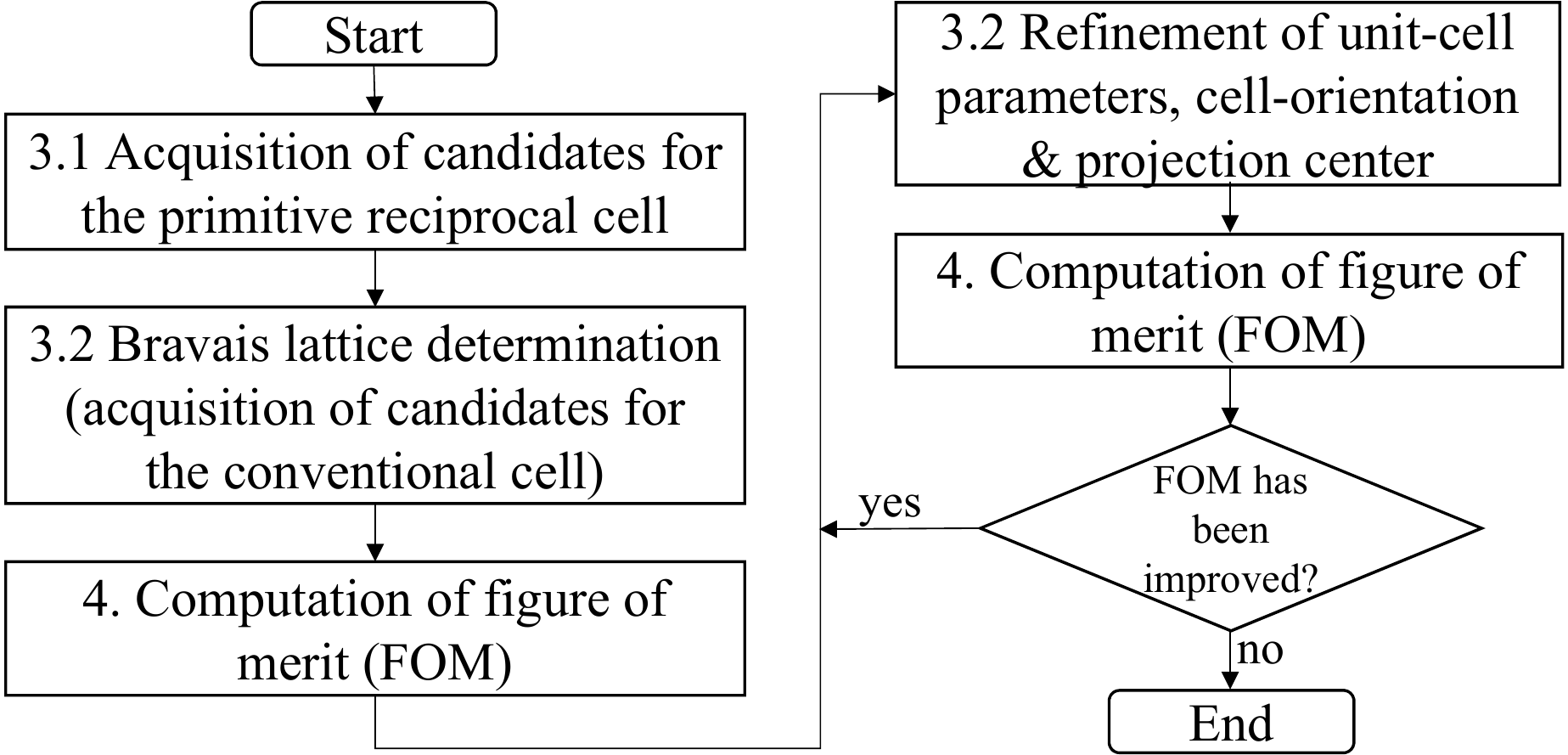}}
\end{center}
\caption{Flowchart of the software; sections~3.1, 3.2, and 4 explain the respective parts.}
\label{fig: Flowchart of the new software}
\end{figure}

This method implemented in the new software, 
is executed between acquisition of the primitive-cell and the calculation of figures of merit (Figure~\ref{fig: Flowchart of the new software}).
%In particular, in order to develop an indexing method, it is not necessary to consider specific Bravais types,
%as long as the influence of systematic absence due to space-group symmetries is discussed.

In the refinement stage, the following parameters are fit to the band positions (and widths, according to the user's choice):
\begin{itemize} 
	\item $s$: scale of the unit cell.
	\item ($\Delta x$, $\Delta y$, $\Delta z$): projection-center shift.  
	\item $\theta^\prime$, $\sigma^\prime$, $\psi^\prime$: Euler angles to represent an orthogonal matrix:
\begin{eqnarray*}
	g(\theta^\prime, \sigma^\prime, \varphi^\prime) &=& 
	\begin{pmatrix}
		\cos \theta^\prime & \sin \theta^\prime & 0 \\
		-\sin \theta^\prime & \cos \theta^\prime & 0 \\
		0 & 0 & 1 \\
	\end{pmatrix} \\
	& & \hspace{-20mm}\times
	\begin{pmatrix}
		1 & 0 & 0 \\
		0 & \cos \sigma^\prime & \sin \sigma^\prime \\
		0 & -\sin \sigma^\prime & \cos \sigma^\prime \\
	\end{pmatrix}
	\begin{pmatrix}
		\cos \varphi^\prime & \sin \varphi^\prime & 0 \\
		-\sin \varphi^\prime & \cos \varphi^\prime & 0 \\
		0 & 0 & 1 \\
	\end{pmatrix}.
\end{eqnarray*}
		
	\item unit-cell parameters represented by the five entries of the lower triangle matrix:
\begin{eqnarray*}
A  :=	\begin{pmatrix}
		1 & 0 & 0 \\
		a_{21} & a_{22} & 0 \\
		a_{31} & a_{32} & a_{33}
	\end{pmatrix}.
\end{eqnarray*}
The above $A$ is obtained by applying the Cholesky decomposition to the following symmetric matrix,
and setting $a^*$ to 1 in order to normalize the scale: 
\begin{eqnarray}\label{eq: matrix A A^T}
A A^T
  &=& \begin{pmatrix}
		(a^*)^2 & a^* b^* \cos \gamma^* & a^*c^* \cos \beta^* \\
		a^* b^* \cos \gamma^* & (b^*)^2 & b^*c^* \cos \alpha^* \\
		a^*c^* \cos \beta^*  & b^*c^* \cos \alpha^* & (c^*)^2
     \end{pmatrix} \nonumber \\
  &=& 
  \begin{pmatrix}
		a^2 & ab \cos \gamma & ac \cos \beta \\
		ab \cos \gamma & b^2 & bc \cos \alpha \\
		ac \cos \beta    & bc \cos \alpha & c^2
     \end{pmatrix}^{-1},
\end{eqnarray}
where $a^*, b^*, c^*, \alpha^*, \beta^*, \gamma^*$ are the reciprocal unit-cell parameters, and
$a, b, c, \alpha, \beta, \gamma$ are the unit-cell parameters.
\end{itemize} 

When a Kikuchi band corresponds to the Miller index $m (hk\ell)$,
the foot $(X^{cal}, Y^{cal})$ of the perpendicular from the pattern center to the Kikuchi band, can be computed by:
\begin{eqnarray}
X^{cal} &=& \frac{-x z}{x^2 + y^2} (1 - \Delta z) + \Delta x, \label{eq: definiton of X^{cal}} \\
Y^{cal} &=& \frac{-y z}{x^2 + y^2} (1 - \Delta z) + \Delta y, \label{eq: definiton of Y^{cal}}
\end{eqnarray}
where $x, y, z$ are the parameters computed by: 
\begin{eqnarray}
\begin{pmatrix} x & y & z \end{pmatrix} &\underset{def}{=}& m \begin{pmatrix} h & k & \ell \end{pmatrix} A g(\theta^\prime, \sigma^\prime, \varphi^\prime).\label{eq: definiton of (x, y, z)}
\end{eqnarray}

$X^{cal}$, $Y^{cal}$ are independent of the choice of $m$.
From Eqs.(\ref{eq: definiton of X^{cal}}), (\ref{eq: definiton of Y^{cal}}),
$\Delta z$ and the scale of the $z$-axis cannot be simultaneously determined only from $(X^{cal}$, $Y^{cal})$.
In order to obtain both, it is necessary to use the bandwidths calculated by:
\begin{eqnarray}\label{eq:band width 2}
\beta^{cal} &=& (\tan(\sigma^{cal} + \theta^{cal}) - \tan(\sigma^{cal} + \theta^{cal}))(1 - \Delta z).
\end{eqnarray}
The values of $\sigma^{cal}$ and the Bragg angle $\theta^{cal}$
are computed by
\begin{eqnarray*}
	\sigma^{cal} &=& \arctan( z / \sqrt{x^2 + y^2} ), \\
	\theta^{cal} &=& \arcsin( s \lambda \sqrt{x^2+y^2+z^2} / 2 ),
\end{eqnarray*}
where $s$ is the scale of the unit cell (required by the above scaling of $A$), and $\lambda$ is the wavelength of the electron beam.
Although the uncertainty in the value of $\lambda$ has been pointed out \cite{Nolze2017a}, 
it can be included in the uncertainty in band widths. Therefore, $\lambda$ is fixed to the input value.

The refinement process is carried out by non-linear least squares method of the Levenberg-Marquardt algorithm. 
The parameters obtained in the indexing process are used as the initial parameters of $s$ and $A$.
The integers $m$ in Eq.(\ref{eq: definiton of (x, y, z)}) are reassigned in every iteration of the refinement process, by checking which $m$ gives the $\beta^{cal}$ closest to the observed $\beta^{obs}$.

\section{De Wolff figures of merit for EBSD indexing}

Some sorting system is required for finding the most plausible one 
from the multiple candidate solutions
in a short time. For the orientation determination of the EBSD patterns, the \textit{Confidence Index} (CI, Field (1997)\nocite{Field97}) based on the number of ``votes'' (Wright(1992)\nocite{Wright92}, the \textit{Fit} based on the difference between the computed bands and the detected bands), and the \textit{Image Quality} (IQ) are used.

Herein, 
use of the idea of the de Wolff figure of merit $M_n$ is proposed. 
Although a number of new figures of merit have been proposed for powder indexing, $M_n$ has been the most efficient indicator in use for long. In particular, it is possible to judge whether or not a plausible solution is included in the output, just by checking the largest value of $M_n$. In Section~\ref{Computational results}, we shall see that its generalizations to 2D and 3D data also have this property.

The de Wolff figure of merit $M_n$ evaluates the similarity between the set of observed $q$-values ($=1/d^2$, $d$: $d$-spacing) $0 < Q_1^{obs} < \ldots < Q_n^{obs}$
and the set of computed $0 < q_1 < \ldots < q_N$ by: 
\begin{eqnarray}
	M_n = \bar{\epsilon} / \delta,
\end{eqnarray}
where $\bar{\epsilon}$ and $\delta$ are the \textit{average discrepancy} and the \textit{actual discrepancy}, respectively, which are defined by:
\begin{eqnarray*}
	\bar{\epsilon} &:=& Q_n^{obs} / 2 N, \label{eq:average discrepancy} \\
	\delta &:=& \frac{1}{n} \sum_{i=1}^n \abs{Q_i^{obs} - Q_i^{cal}}, \label{eq:actual discrepancy} \\
	& & \hspace{-10mm} Q_i^{cal}: \text{computed $q$-value closest to the observed } Q_i^{obs}. \nonumber
\end{eqnarray*}

When it is assumed that $Q_i^{obs}$ ($i = 1, \ldots, n-1$) and $q_i$ are uniformly distributed in the interval $[0, Q_n^{obs}]$,
the $\bar{\epsilon}$ in Eq.(\ref{eq:average discrepancy}) is an approximation of the mean value of $\delta$
(Wu (1988)\nocite{Wu88}). Namely, the expected value of the shortest distance between $Q$ and $q_1, \ldots, q_n$ equals:
$$
	\bar{\epsilon} \approx E \left[ \min_{i = 1, \ldots, N} \{ \abs{ Q - q_i } \} \right].
$$

In Appendix~C, this idea is extended to data of general dimensions. In particular, the obtained figures of merit is scale-free, similarly to the original $M_n$.

If a set of computed points $x_1, \ldots x_N$ and $X$ are uniformly distributed in an $s$-dimensional hypersphere of radius $R$, the expected value of the shortest distance between $\{ x_1, \ldots, x_N \}$ and $X$ can be approximated by the following asymptotic formula:
\begin{eqnarray*}
E \left[ \min_{i = 1, \ldots, N} \{ \abs{ X - x_i } \} \right] \sim \frac{\Gamma(1/s)}{s} \frac{R}{ N^{1/s} },
\end{eqnarray*}
where $\Gamma(z)$ is the Gamma function $\int_{0}^\infty t^{z-1} e^{-t} dt$.
By using the volume $V = (\sqrt{\pi} R)^s / (\Gamma(s/2+1) N)$ of the $s$-dimensional hypersphere, 
the following is obtained:
\begin{eqnarray}\label{eq: formula for general convex bodies}
E \left[ \min_{i = 1, \ldots, N} \{ \abs{ X - x_i } \} \right]
\sim
\frac{\Gamma(s/2+1)^{1/s} \Gamma(1/s)}{ \sqrt{\pi} s} \left( V/N \right)^{1/s} .
\end{eqnarray}
 
For any point configuration in a convex body of volume $V$, Eq.(\ref{eq: formula for general convex bodies}) holds, because the influence of the boundary can be ignored for sufficiently large $N$.

In particular, the formulas for the dimensions $s = 2, 3$ are:
%\begin{description}
%\item[($2D$ ball of radius $R$)]
%\begin{eqnarray*}
%E \left[ \min_{1 \leq i \leq N} \{ \abs{ X - x_i } \} \right]
%\sim \frac{\sqrt{\pi}}{2} \frac{R}{\sqrt{N}}.
%\end{eqnarray*}
%
%\item[($3D$ ball of radius $R$)]
%\begin{eqnarray*}
%E \left[ \min_{1 \leq i \leq N} \{ \abs{ X - x_i } \} \right]
%&\sim& 
%\frac{\Gamma(1/3)}{3}
%\frac{R}{N^{1/3}}.
%\end{eqnarray*}
%
%\end{description}
\begin{description}
\item[(Case of $2D$ objects of volume $V$)]

\begin{eqnarray}\label{eq: formula for 2D objects}
E \left[ \min_{i = 1, \ldots, N} \{ \abs{ X - x_i } \} \right]
\sim \frac{1}{2} (V / N)^{1/2}. \label{eq:2D average discrepancy}
\end{eqnarray}

\item[(Case of $3D$ objects of volume $V$)]

\begin{eqnarray}\label{eq: formula for 3D objects}
E \left[ \min_{i = 1, \ldots, N} \{ \abs{ X - x_i } \} \right]
&\sim&
\Gamma(1/3) \left( \frac{V}{36 \pi N} \right)^{1/3} \\
&\approx&
2.6789 \left( \frac{V}{36 \pi N} \right)^{1/3}. \nonumber
\end{eqnarray}

\end{description}

In the following examples, 
it is explained how Eqs.(\ref{eq: formula for 2D objects}) and (\ref{eq: formula for 3D objects}) can be used for orientation determination and ab-initio indexing.

\begin{description}
\item[Example 1](For comparison of band positions)
\end{description}
\vspace{-2mm}
The feet $P_i^{obs} = (X_i^{obs}, Y_i^{obs})$ ($i = 1, \ldots, n$) of the perpendiculars from the pattern center to the Kikuchi bands, can be regarded as a set of coordinates distributed in a 2D circle of radius $R := \max_{i = 1, \ldots, n} \{ (X_i^{obs})^2 + (Y_i^{obs})^2 \}$.
If $(X_j^{cal}, Y_j^{cal})$ ($j = 1, \ldots, N$) are the computed band positions in the circle, these two sets can be compared 
by setting $V$ in Eq.(\ref{eq:2D average discrepancy}) to the area $V = \pi R^2$ of the circle; the ratio $M_{n, N} = \bar{\epsilon} / \delta$ is computed by:
\begin{eqnarray*}
	\bar{\epsilon} &:=& \frac{R}{2} \sqrt{ \frac{\pi}{N} }, \\
	\delta &:=& \frac{1}{n} \sum_{i=1}^n \abs{P_i^{obs} - P_i^{cal}},
\end{eqnarray*}
where $P_i^{cal} = (X_i^{cal}, Y_i^{cal})$ is the computed point closest to $P_i^{obs}$.

\begin{description}
\item[Example 2] 
(For comparison of band positions and widths)
\end{description}
\vspace{-2mm}
The 3D coordinates representing the Kikuchi bands are obtained by considering the bandwidths $\beta_i^{obs}$ 
as the third coordinates:
\begin{itemize}
\item ${\mathcal P}_i^{obs} = (X_i^{obs}, Y_i^{obs}, \beta_i^{obs})$ ($i = 1, \ldots, n$),

\item $(X_j^{cal}, Y_j^{cal}, \beta^{cal}_j)$ ($j = 1, \ldots, N$),
where $\beta^{cal}_j$ is approximated by using Eq.(\ref{eq:band width 2}) and the following:
\begin{eqnarray*}
\sigma^{cal}  &=& \arctan(z / \sqrt{x^2 + y^2}), \\
2 \theta^{cal}  &\approx& 2 \sin \theta^{cal}  = s \lambda \sqrt{x^2 + y^2 + z^2}.
\end{eqnarray*}
%The approximation $\approx$ is frequently used to obtain the Bragg angle $\theta$ from the bandwidths.

\end{itemize}

The above ${\mathcal P}_i^{obs}$ are distributed in the cylinder
with the radius $R = \max_{i = 1, \ldots, n} \{ (X_i^{obs})^2 + (Y_i^{obs})^2 \}$ and the height $h = \max_{i = 1, \ldots, n} \{ \beta_i^{obs} \}$.
Therefore, $V$ in Eq.(\ref{eq: formula for 3D objects}) is set to $\pi R^2 h$. The figure of merit $M_{n, N} = \bar{\epsilon} / \delta$ is computed by:
\begin{eqnarray*}
	\bar{\epsilon} &:=& 
	 \Gamma(1/3) \left( \frac{ R^2 h }{ 36 N } \right)^{1/3}, \\
	\delta &:=& \frac{1}{n} \sum_{i=1}^n \abs{ {\mathcal P}_i^{obs} - {\mathcal P}_i^{cal}}.
\end{eqnarray*}
where ${\mathcal P}_i^{cal} = (X_i^{cal}, Y_i^{cal}, \beta_i^{cal})$ is the computed point closest to ${\mathcal P}_i^{obs}$.

\vspace{2mm}
In the definition, % of the 3D version, it is possible to adjust the influence of the bandwidths by changing the common scale of $\beta^{obs}_i$ and $\beta^{cal}_i$.
the number of computed points $N$ is also a parameter, because infinitely many non-visible bands are theoretically included in the range of the observed EBSD pattern.
This $N$ can be automatically determined, by setting the upper thresholds for the $d$-values and imposing the following constraint on the  generated Miller indices.
\begin{itemize}
\item for each indexing solution, the upper threshold for the $d$-values of the computed bands
is set to the minimum value required to assign a Miller index to every observed band.

\item In $(X_i^{cal}, Y_i^{cal})$ ($i = 1, \ldots, N$), overlapping bands $m ( h k \ell )$ ($m \ne 0$: integer) 
appear only once. Namely, only the narrowest bandwidths $\beta_j^{cal}$ are contained and compared.

\end{itemize}
By doing the above,  
solutions that can assign $h k \ell$ with smaller $\abs{h}, \abs{k}, \abs{\ell}$ to the bands, obtain larger $M_{n, N}$.
This is a heuristic for ranking the true solution above the derivative lattices.
Very flat or thin unit cells, are less likely to be selected consequently.

\section{Computational results and discussion}
\label{Computational results}

The proposed method was implemented using C++,
and applied to the analysis of dynamically simulated Kikuchi patterns (Figures~1--4 in Appendix~B) and experimental patterns (Figures~\ref{fig: Band positions and widths extracted from an experimental pattern of Cemetite}--\ref{fig: Band positions and widths extracted from an experimental pattern of Silico Ferrite}).
The program was run on an Intel Core i7-5930k CPU (3.50 GHz) without parallel computation. 
The results are presented in Tables~\ref{The optimum solutions when band widths are not used}~and~\ref{The optimum solutions when band widths are used}.

The quick search and exhaustive search were carried out by using the algorithm in Table~\ref{The algorithm for ab-initio EBSD indexing (without information about $d$-spacings}.
The only difference between them is that the check (*) in the footnote of Table~\ref{The algorithm for ab-initio EBSD indexing (without information about $d$-spacings} is performed in the former search.
The search parameters commonly used for the test, are listed in Table~1 of Appendix~A. 

The simulated patterns were created by using the Bruker's commercial software \textit{DynamicS} \cite{Winkelmann2007}. 
The parameters used for the simulation are presented in Table~1 of Appendix~B.
In particular, the used coordinates of the projection centers were exact.
%The acceleration voltage of electron beam was set as 20 kV.

The experimental EBSD patterns were prepared
by using an SEM-EBSD system (JEOL 7001F-EDAX DigiView camera)
with the $20 kV$ electron-accelerating voltage and the beam current up to $14 nA$. 
Hence, the wavelength of the electron beam is $8.588510^{-12} m$ (relativistic effects are considered).
Precise projection centers were also available for these experimental data, 
since the second author used the pattern matching technique to obtain them (see Nolze \textit{et.\ al.} (2017)\nocite{Nolze2017b}; this technique requires pattern simulation based on the phase information).
Therefore, 
the results for imprecise projection centers are also presented, after the results for the precise projection centers.

%In addition, the error in projection center can be also an indirect reason for 
%the ambiguity caused by inaccurate band edges.
For the test, 
the band slopes, its perpendiculars and band widths were manually extracted (without checking band profiles),
because the existing automatic methods could not obtain satisfactory results.
In order to obtain a reliable indexing result, at least more than a dozen of band coordinates should be prepared.
However, software development for extracting many band edges is not the scope of this article.
More precisely, visually recognizable dark lines surrounding a bright band were used as the band edges.

In band searching without assuming the band-profile model, 
it is basically impossible to estimate precise errors of 
the parameters $\varphi$, $\sigma$ to represent the band coordinates (Figure~\ref{fig: transformation_of_EBSD_pattern}) 
and $\sigma \pm \theta$ to represent the band widths (Eq.(\ref{eq:band width})). 
Therefore, all the error values of the input angles are estimated to 1~degree, and used to calculate the propagation errors of the unit cell parameters.
The error does not seem to be overestimation, considering the obtained propagation errors (Tables~\ref{The optimum solutions when band widths are not used}--\ref{The optimum solutions when band widths are used}). 
In particular, large errors in the band widths are justified by the uncertainty of the wavelength due to the energy loss of the electron, and the ambiguity in the definition of the band width. Comparing to these factors, the influence of the accelerating voltage error is small.

In Figures~\ref{fig: Band positions and widths extracted from an experimental pattern of Cemetite},
\ref{fig: Band positions and widths extracted from an experimental pattern of Silico Ferrite}, 
the yellow lines present the band slopes used to determine the band edges. 
The band positions are the midpoints of the band edges if the spherical coordinates $(\sin \sigma \cos \varphi, \sin \sigma \sin \varphi, \cos \sigma)$ are used.

%The software also refines the projection center, although there is ambiguity in determination of $\Delta_z$, as mentioned in the paragraph following Eq.(\ref{eq: definiton of (x, y, z)}).
% (see each figures for the used band information).

\begin{figure}[htbp]
\begin{center}
\scalebox{0.3}{\includegraphics{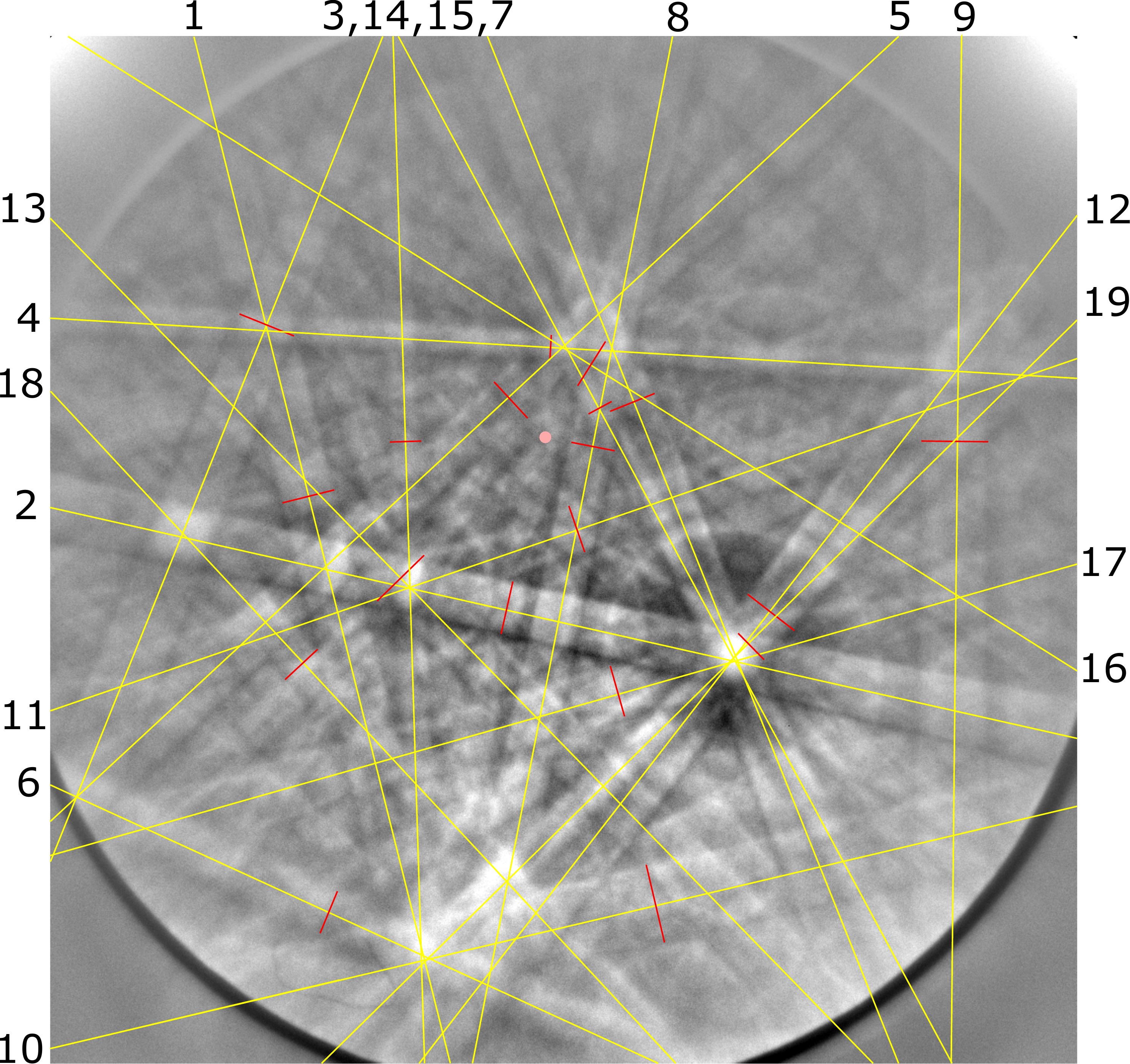}}
\end{center}
\caption{Band slopes and widths extracted from an experimental pattern of Cementite ($1040 \times 1040$ $px^2$)}
\label{fig: Band positions and widths extracted from an experimental pattern of Cemetite}
\end{figure}

\begin{figure}[htbp]
\begin{center}
\scalebox{0.3}{\includegraphics{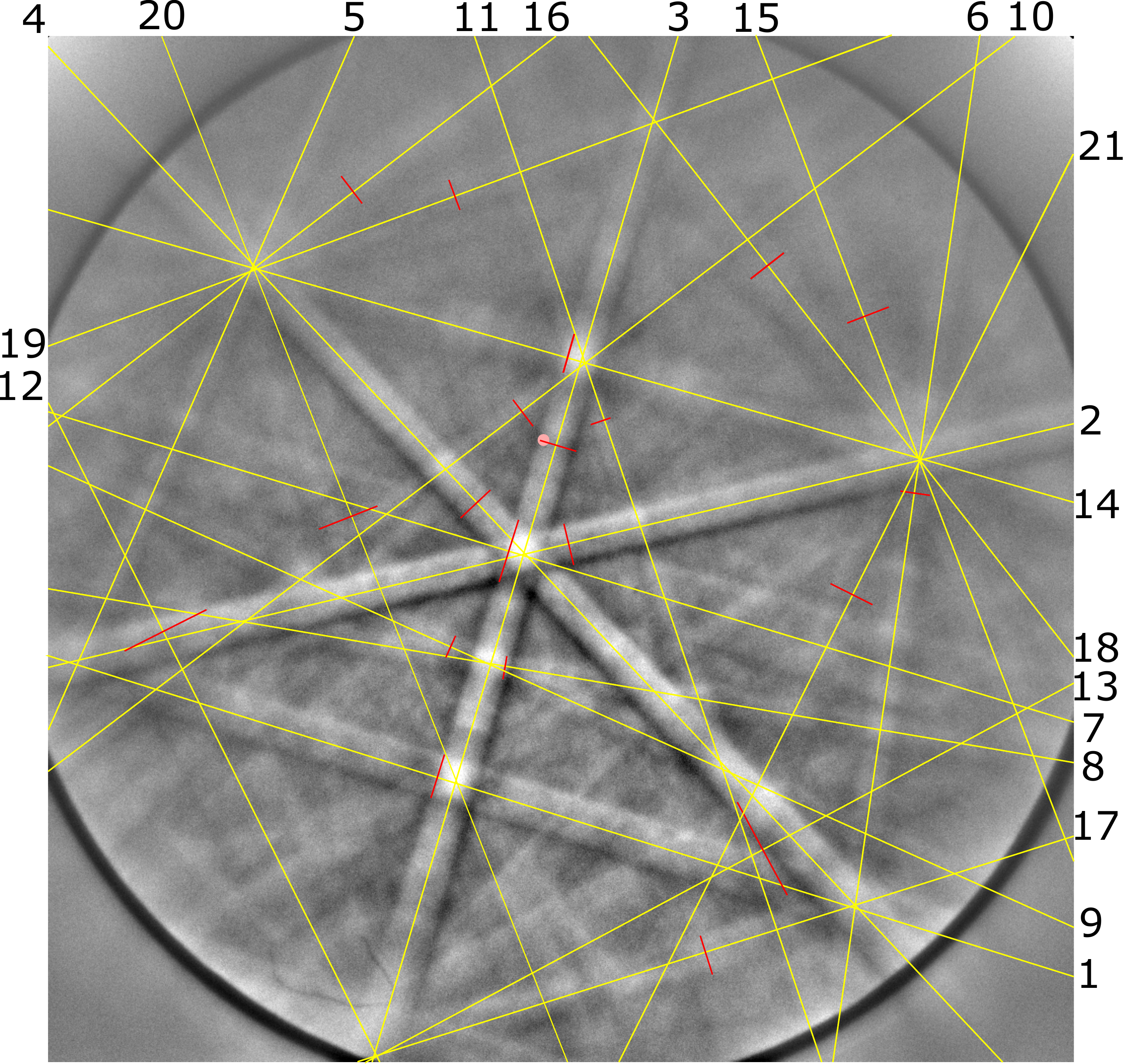}}
\end{center}
\caption{Band slopes and widths extracted from an experimental pattern of Silico Ferrite of $Ca$ \& $Al$  ($1040 \times 1040$ $px^2$)}
\label{fig: Band positions and widths extracted from an experimental pattern of Silico Ferrite}
\end{figure}

When precise projection centers were used, the program succeeded in acquisition of the correct cell, except for the two triclinic cases among 6 test data. The new figures of merit ranked it at the top among the solutions of the same Bravais type. 
However, the scales of the obtained unit-cell parameters are 2.1--8.9 \% smaller than the literature values.

Owing to the inaccuracy of the used band widths,
the $M_{n, N}$ values in Table~\ref{The optimum solutions when band widths are used} are smaller than those in Table~\ref{The optimum solutions when band widths are not used}. 
In particular, rather small $M_{n, N} = 4.33$ was obtained for the cementite sample when the bandwidths were used, although 
the solution was almost identical to the case without band widths.
%With regard to the Silico-ferrite sample,
%the difference from the literature values was increased by using the bandwidths.
%In this case, the narrowest bandwidths may not have been detected well from the experimental pattern.

\onecolumn
\begin{landscape}
\begin{table}[htbp]
\caption{Comparison of the optimal indexing solutions and the parameters in the literature (when bandwidths are not used$^a$)}
\label{The optimum solutions when band widths are not used}
\begin{minipage}{\textwidth}
\begin{footnotesize}
\begin{tabular}{lllrrrrr%r
}      % Alignment for each cell: l=left, c=center, r=right
\hline
\begin{tabular}{l}
Number of \\ used bands
\end{tabular}
&
\begin{tabular}{l}
Success/Failure in \\ quick search \\ (time (sec.), $M_{n, N}$)
\end{tabular} &
\begin{tabular}{l}
Success/Failure in \\ exhaustive search \\ (time (sec.), $M_{n, N}$)
\end{tabular} &
$a/c$ (error$^b$)    & $b/c$ (error$^b$)     & $\alpha$(deg.) (error$^b$)    & $\beta$(deg.) (error$^b$)     & $\gamma$(deg.) (error$^b$) %& Bravais type 
\\
\hline
\multicolumn{8}{l}{{\bf Simulated patterns:}} \\
\multicolumn{8}{l}{{\bf Ni} (cubic(F), $a/c=b/c=1$):} \\
20 & S (7.40, $M_{20, 41} = 46.60$) & S (243.83, $M_{20, 41} = 47.56$)  & 1       & 1        & 90       & 90         & 90 %& cubic (F)   
\\
\multicolumn{8}{l}{{\bf Fe} (cubic(I), $a/c=b/c=1$):} \\
23 & S (11.78, $M_{23, 57} = 57.73$)   & S (426.62, $M_{23, 58} = 59.03$)  & 1       & 1        & 90       & 90         & 90 %& cubic (I) 
\\
\multicolumn{8}{l}{{\bf Zn} (hexagonal, $a/c=b/c = 0.5387$):} \\
23 & S (3.17, $M_{23, 107} = 39.47$) & S (144.68, $M_{23, 105} = 39.85$) & 0.540 (0.008)        & 0.540 (0.008)        & 90       & 90         & 120    %& hexagonal 
\\
\multicolumn{8}{l}{{{\bf Silico-ferrite of calcium and aluminum}} (triclinic, $a/c = 0.881$, $b/c = 0.897$, 
$\alpha = 94.11$, $\beta = 111.4$, $\gamma = 110.3$):} \\
22 & F (3.26) & F$^c$ (339.39, $M_{22, 301}=21.73$) & 0.589 (0.007)  & 0.588 (0.009)  & 89.82 (0.78)  & 106.99 (1.01) & 119.95 (0.83) %& triclinic  
\\ \\
\multicolumn{8}{l}{{\bf Experimental patterns:}} \\
\multicolumn{8}{l}{{\bf Spheroidal cementite} (orthorhombic (P), $a/c=0.6711$, $b/c = 0.7546$):} \\
19 &  F (0.91) & S (20.03, $M_{19, 271} = 16.73$) & 0.663 (0.011)  & 0.745 (0.012)        & 90       & 90         & 90 %& orthorhombic (P)  
\\
\multicolumn{8}{l}{{{\bf Silico-ferrite of calcium and aluminum}} (triclinic, $a/c = 0.881$, $b/c = 0.897$, 
$\alpha = 94.11$, $\beta = 111.4$, $\gamma = 110.3$):} \\
21 & F (0.77) & F$^c$ (53.67, $M_{21, 347}=7.75$) & 0.764 (0.013)  & 0.829 (0.017)  & 91.37 (1.13)  & 100.36 (1.04) & 102.29 (0.76) %& triclinic  
\\
\hline
\end{tabular}
\end{footnotesize}
	\footnotetext[1]{
The pattern-center shift $\Delta z$ in the direction perpendicular to the screen was fixed to 0 for low-symmetric cells, considering the ambiguity pointed out by Alkorta (2013)\nocite{Alkorta2013}.
	}
	\footnotetext[2]{
These error values are computed as the propagation errors of the estimated error 1~degree of 
the input angles $\sigma$, $\varphi$ and $\sigma \pm \theta$ (see Figure~\ref{fig: transformation_of_EBSD_pattern}
and Eq.(\ref{eq:band width})).
	}
	\footnotetext[3]{
These triclinic cases are regarded as failures, since the differences in the length-ratios and angles from the literature values exceeded 10\%, although all the bands were indexed.
For the simulated pattern, as seen from $3a : 3b : 2c=0.884 : 0.881 : 1$, 
the parameters
that attained the largest $M_{n, N}$ value among triclinic solutions,
are close to those of the derivative lattice of the true solution. 
	}
\end{minipage}
\end{table}

\end{landscape}
\begin{landscape}

\begin{table}[htbp]
\caption{Comparison of the optimal indexing solutions and the parameters in the literature when bandwidths are used}
\label{The optimum solutions when band widths are used}
\begin{minipage}{\textwidth}
\begin{footnotesize}
\begin{tabular}{lllrrrrrr%r
} % Alignment for each cell: l=left, c=center, r=right
\hline
\begin{tabular}{l}
Number of \\ used bands
\end{tabular}
&
\begin{tabular}{l}
Success/Failure in \\ quick search \\ (time (sec.), $M_{n, N}$)
\end{tabular} &
\begin{tabular}{l}
Success/Failure in \\ exhaustive search \\ (time (sec.), $M_{n, N}$)
\end{tabular}
& $a$($\AA$) (error$^b$) & $b$($\AA$) (error$^b$) & $c$($\AA$) (error$^b$) & $\alpha$(deg.) & $\beta$(deg.) & $\gamma$(deg.) %& Bravais type
 \\
\hline
\multicolumn{9}{l}{{\bf Simulated patterns:}} \\
\multicolumn{9}{l}{{\bf Ni} (cubic(F), $a=b=c=3.516 (\AA)$):} \\
20 & S (6.15, $M_{20, 41} = 22.62$) & S (243.13, $M_{20, 39} = 23.00$)    & 3.397 (0.235) & 3.397 (0.235)  & 3.397 (0.235)    & 90       & 90         & 90 %& cubic (F)    
\\
\multicolumn{9}{l}{{\bf Fe} (cubic(I), $a=b=c=2.866 (\AA)$):} \\
23 & S (13.82, $M_{23, 57} = 37.78$)  & S (426.15, $M_{23, 57} = 38.13$) & 2.805 (0.184) & 2.805 (0.184)  & 2.805 (0.184)  & 90       & 90         & 90 %& cubic (I)    
\\
\multicolumn{9}{l}{{\bf Zn} (hexagonal, $a=b=2.665$, $c=4.947 (\AA)$):} \\
23 & S (2.63, $M_{23, 103} = 24.29$) & S (144.41, $M_{23, 103} = 24.29$)  & 2.567 (0.179)  & 2.567 (0.179)   & 4.706 (0.344)  & 90       & 90         & 120     %& hexagonal 
\\
\multicolumn{9}{l}{\bf Silico-ferrite of calcium and aluminum} \\
\multicolumn{9}{l}{(triclinic, $a=10.40$, $b = 10.59$, $c = 11.81 (\AA)$, 
$\alpha = 94.11$, $\beta = 111.4$, $\gamma = 110.3$(deg.)$^\S$:} \\
22 & F (2.27)  & F$^c$ (289.03) &         &         & &        &          &  %& triclinic   
\\ \\
\multicolumn{9}{l}{{\bf Experimental patterns:}} \\
\multicolumn{9}{l}{{\bf Spheroidal cementite}
(orthorhombic (P), $a=4.526$, $b = 5.089$, $c = 6.744 (\AA)$ \cite{Gardin63}):} \\
19 & F (0.25) & S (20.14, $M_{19, 262} = 4.33$)  & 4.122 (0.330)    &   4.659 (0.4121)    & 6.245 (0.5411)  & 90       & 90         & 90 %& orthorhombic (P)  
\\
\multicolumn{9}{l}{\bf Silico-ferrite of calcium and aluminum} \\
\multicolumn{9}{l}{(triclinic, $a=10.40$, $b = 10.59$, $c = 11.81 (\AA)$, 
$\alpha = 94.11$, $\beta = 111.4$, $\gamma = 110.3$(deg.)$^d$:} \\
21 & F (1.01)  & F$^c$ (48.49) &         &         & &        &          &  %& triclinic   
\\
\hline
\end{tabular}
\end{footnotesize}
	\footnotetext[2]{
See the footnote of Table~\ref{The optimum solutions when band widths are not used}.
	}
	\footnotetext[3]{
For these triclinic cases, none of the obtained solutions was close to the true solution.
However, for the simulated pattern, 
a triclinic solution attained $M_{22, 167} = 8.82$ and all the input bands were indexed. 
	}
	\footnotetext[4]{
The unit-cell parameters were obtained by the Rietveld refinement of X-ray diffraction data
(a little distinct from the literature values in Takayama \textit{et. al.} (2018)\nocite{Takayama2018}, owing to the different composition)
}
\end{minipage}
\end{table}

\end{landscape}
\begin{landscape}

\begin{table}[htbp]
\caption{Influence of the projection center shift $(\Delta x, \Delta y, \Delta z)$, when bandwidths are not used; 
results for the 343 cases given by $\Delta x/z, \Delta y/z, \Delta z/z = 0, \pm 0.005, \pm 0.01$ or $\pm 0.02$ ($z$: camera length)}
\label{Results when band widths are not used}
\begin{minipage}{\textwidth}
\begin{tabular}{lccccccr}      % Alignment for each cell: l=left, c=center, r=right
\hline
	& Number of failed cases & \multicolumn{6}{c}{Range of obtained solutions} \\
\begin{tabular}{l}
Used search \\
method 
\end{tabular} &
\begin{tabular}{l}
	($+$ number of failed cases \\
	 in Bravais lattice determination)
\end{tabular} &
$a/c$  & $b/c$ & $\alpha$(deg.) & $\beta$(deg.)  & $\gamma$(deg.) & $M_{n, N}$
\\
\hline
\multicolumn{7}{l}{{\bf Ni} (simulated, cubic(F), $a/c=b/c=1$):} \\
Quick search & $0$ ($+0$) / 343 & 1       & 1        & 90       & 90         & 90 & 4.79--50.98  \\
\multicolumn{7}{l}{{\bf Fe} (simulated, cubic(I), $a/c=b/c=1$):} \\
Quick search & $0$ ($+1$) / 343  & 1       & 1        & \multicolumn{3}{c}{90--90.47} & 5.63--58.85
\\
\multicolumn{7}{l}{{\bf Zn} (simulated, hexagonal, $a/c=b/c = 0.5387$):} \\
Quick search & $0$ ($+0$) / 343  & \multicolumn{2}{c}{0.526--0.555}    & 90       & 90         & 120 & 5.86--45.36 \\
\multicolumn{7}{l}{{\bf Spheroidal cementite} (experimental, orthorhombic (P), $a/c=0.6711$, $b/c = 0.7546$):} \\
Exhaustive search &  $0$ ($+0$) / 343  & 
0.643--0.684	
  & 0.734--0.757   & 90       & 90         & 90 & 8.64--21.22 \\
\hline
\end{tabular}
\end{minipage}
\end{table}

\begin{table}[htbp]
\caption{Influence of the projection center shift $(\Delta x, \Delta y, \Delta z)$, when bandwidths are used;
results for the 343 cases given by $\Delta x/z, \Delta y/z, \Delta z/z = 0, \pm 0.005, \pm 0.01$ or $\pm 0.02$ ($z$: camera length).}
\label{Results when band widths are used}
\begin{minipage}{\textwidth}
\begin{tabular}{lcccccccr}      % Alignment for each cell: l=left, c=center, r=right
\hline
	& Number of failed cases & \multicolumn{6}{c}{Range of obtained solutions} \\
\begin{tabular}{l}
Used search \\
method 
\end{tabular} &
\begin{tabular}{l}
	($+$ number of failed cases \\
	 in Bravais lattice determination)
\end{tabular} &
$a$ ($\AA$) & $b$ ($\AA$) & $c$ ($\AA$) & $\alpha$(deg.) & $\beta$(deg.)  & $\gamma$(deg.) & $M_{n, N}$
 \\
\hline
\multicolumn{8}{l}{{\bf Ni} (simulated, cubic(F), $a=b=c=3.516 (\AA)$):} \\
Quick search & $0$ ($+0$) / 343  & \multicolumn{3}{c}{3.228--3.535}
  & 90       & 90         & 90 & 14.94--23.69
 \\
\multicolumn{8}{l}{{\bf Fe} (simulated, cubic(I), $a=b=c=2.866 (\AA)$):} \\
Quick search & $0$ ($+0$)  / 343 & \multicolumn{3}{c}{2.724--2.962}  & 90       & 90         & 90 & 4.63--38.09 \\
\multicolumn{8}{l}{{\bf Zn} (simulated, hexagonal, $a=b=2.665$, $c=4.947 (\AA)$):} \\
Quick search & $0$ ($+28$) / 343 & \multicolumn{2}{c}{2.405--2.718}
   & 4.468--5.051
 & \multicolumn{2}{c}{87.88--92.51}
       & 118.62--120
 & 11.27--27.48 \\
\multicolumn{8}{l}{{\bf Spheroidal cementite}
(experimental, orthorhombic (P), $a=4.526$, $b = 5.089$, $c = 6.744 (\AA)$; Gardin (1962)\nocite{Gardin63}):} \\
Exhaustive search &  $5$ ($+15$) / 343 & 3.86--4.36    &   4.29--4.90
    & 5.74--6.60
 & 90       & 89.22--93.04
        & 90 
& 3.32--4.50 \\
\hline
\end{tabular}
\end{minipage}
\end{table}

\end{landscape}

%\twocolumn

For each sample with the exception of the failed triclinic case, we prepared $7^3 = 343$ different sets of band coordinates,
by perturbing the projection center.
The shifts used to make the input $\varphi, \sigma$ are as follows:
$$
	\frac{\Delta x}{z}, \frac{\Delta y}{z}, \frac{\Delta z}{z} = 0, \pm 0.005, \pm 0.01, \pm 0.02. \quad \text{($z$: camera length)}
$$ 

\begin{figure}[htbp]
\begin{center}
\begin{tabular}{ll}
(i) & (ii) \\
\scalebox{0.35}{\includegraphics{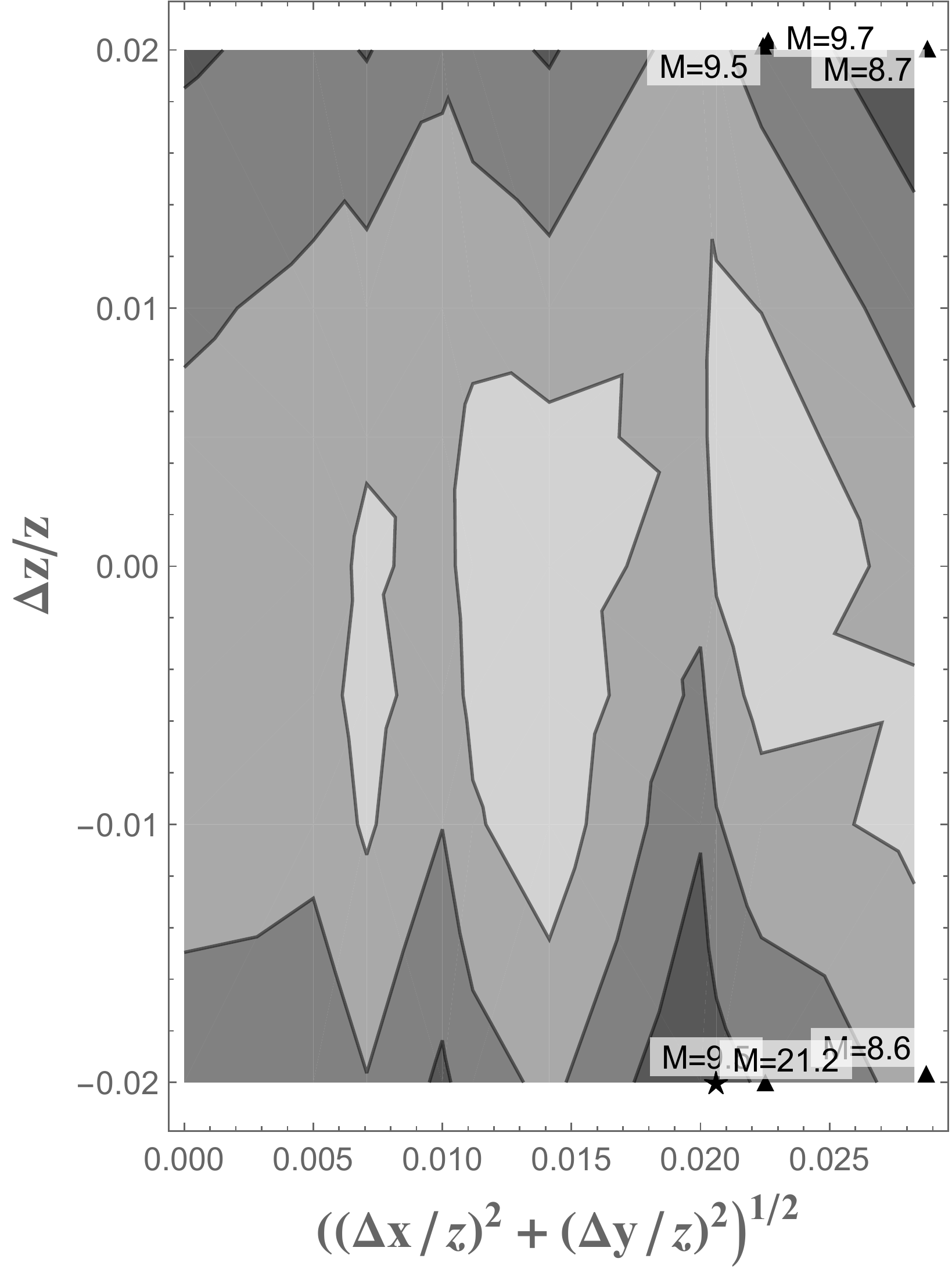}}
\scalebox{0.32}{\includegraphics{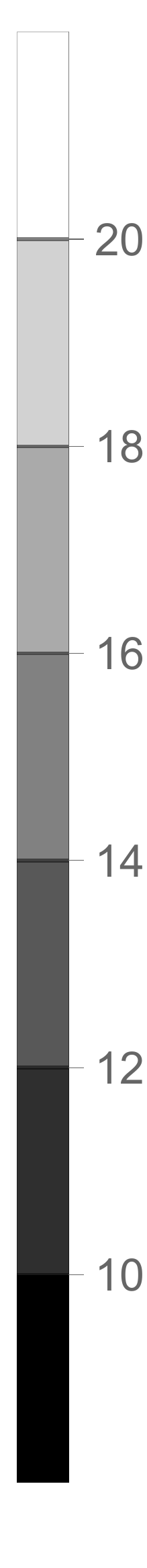}}
 &
\scalebox{0.35}{\includegraphics{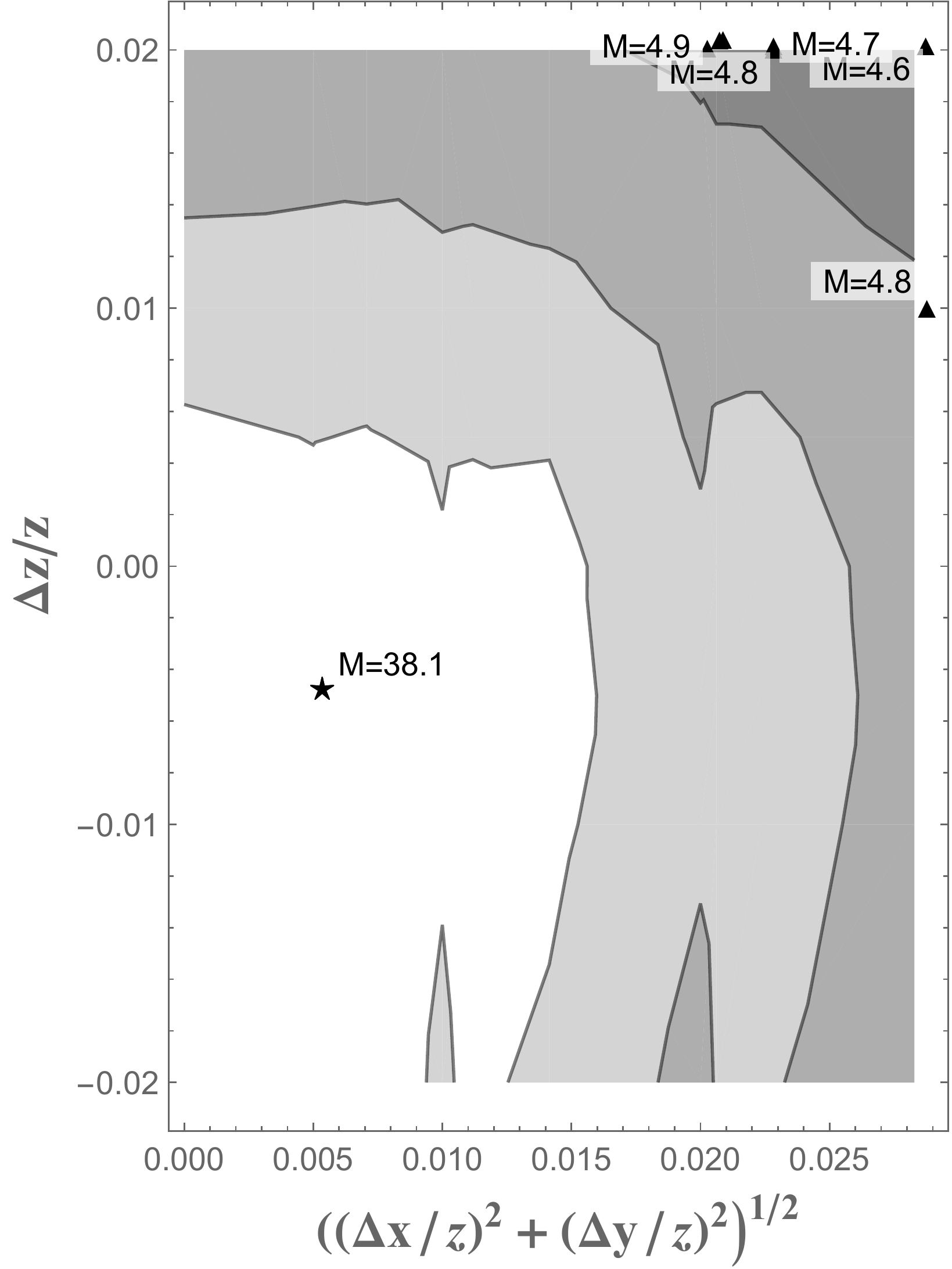}}
\scalebox{0.4}{\includegraphics{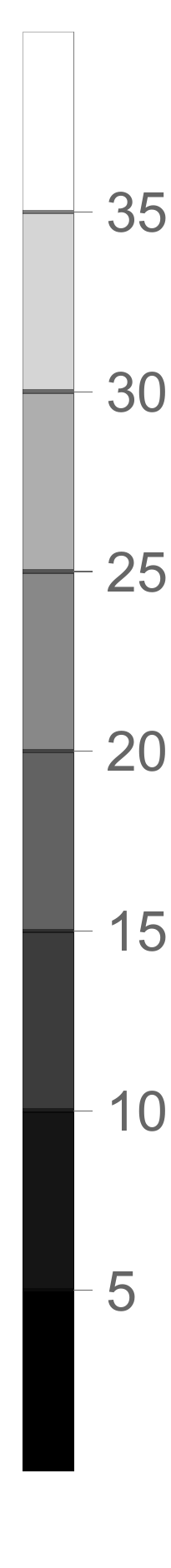}} \\
(iii) & (iv) \\
\scalebox{0.36}{\includegraphics{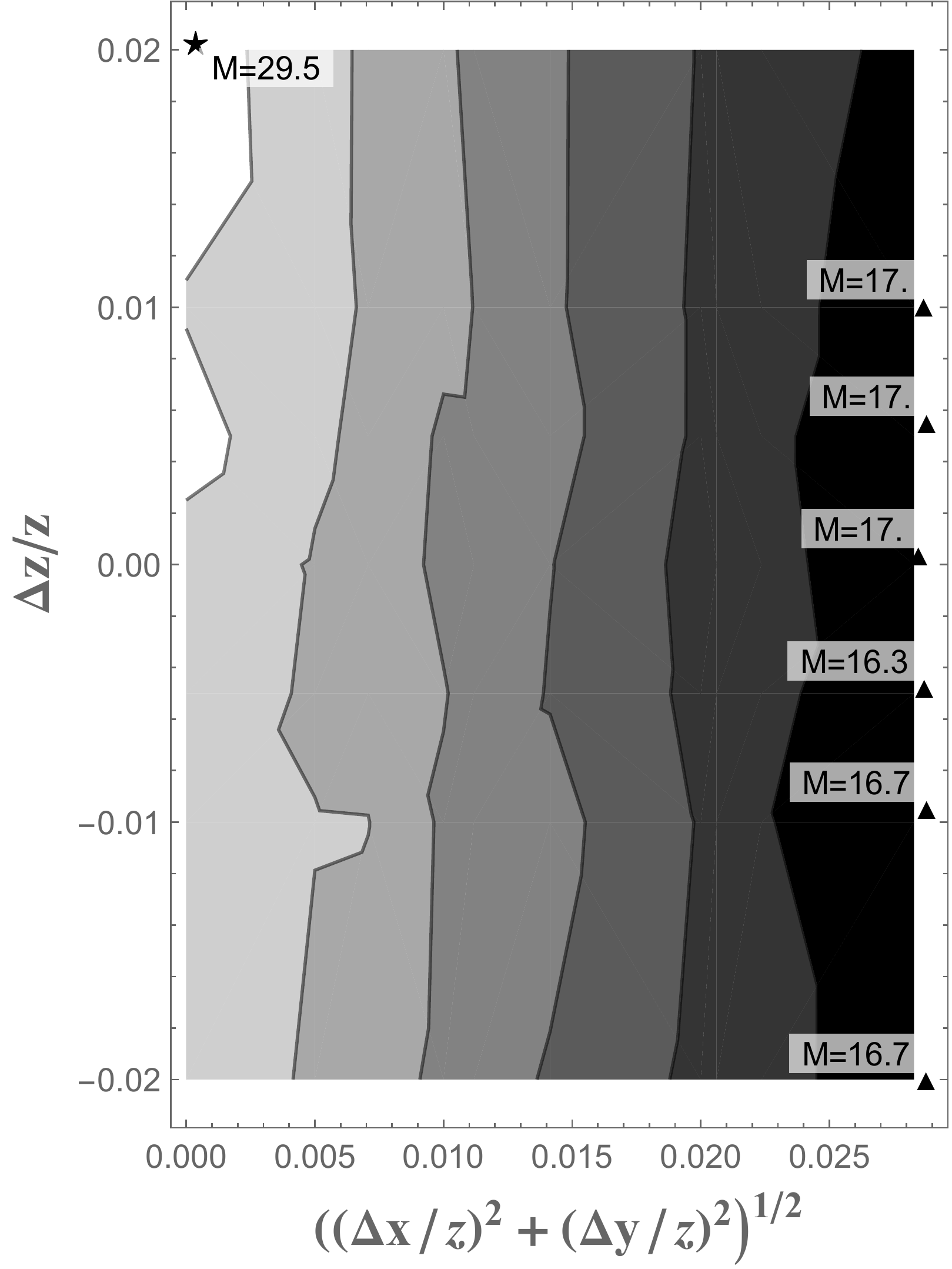}}
\scalebox{0.32}{\includegraphics{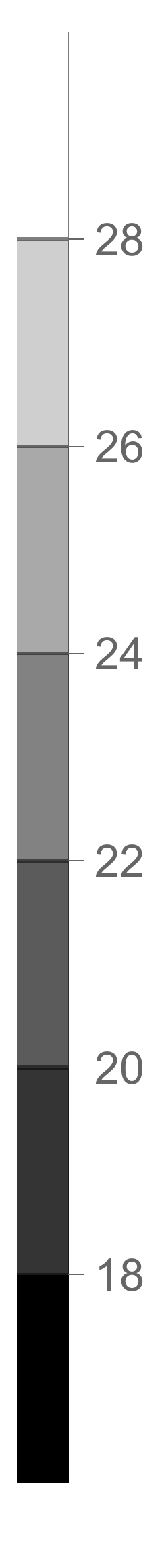}}
 &
\scalebox{0.35}{\includegraphics{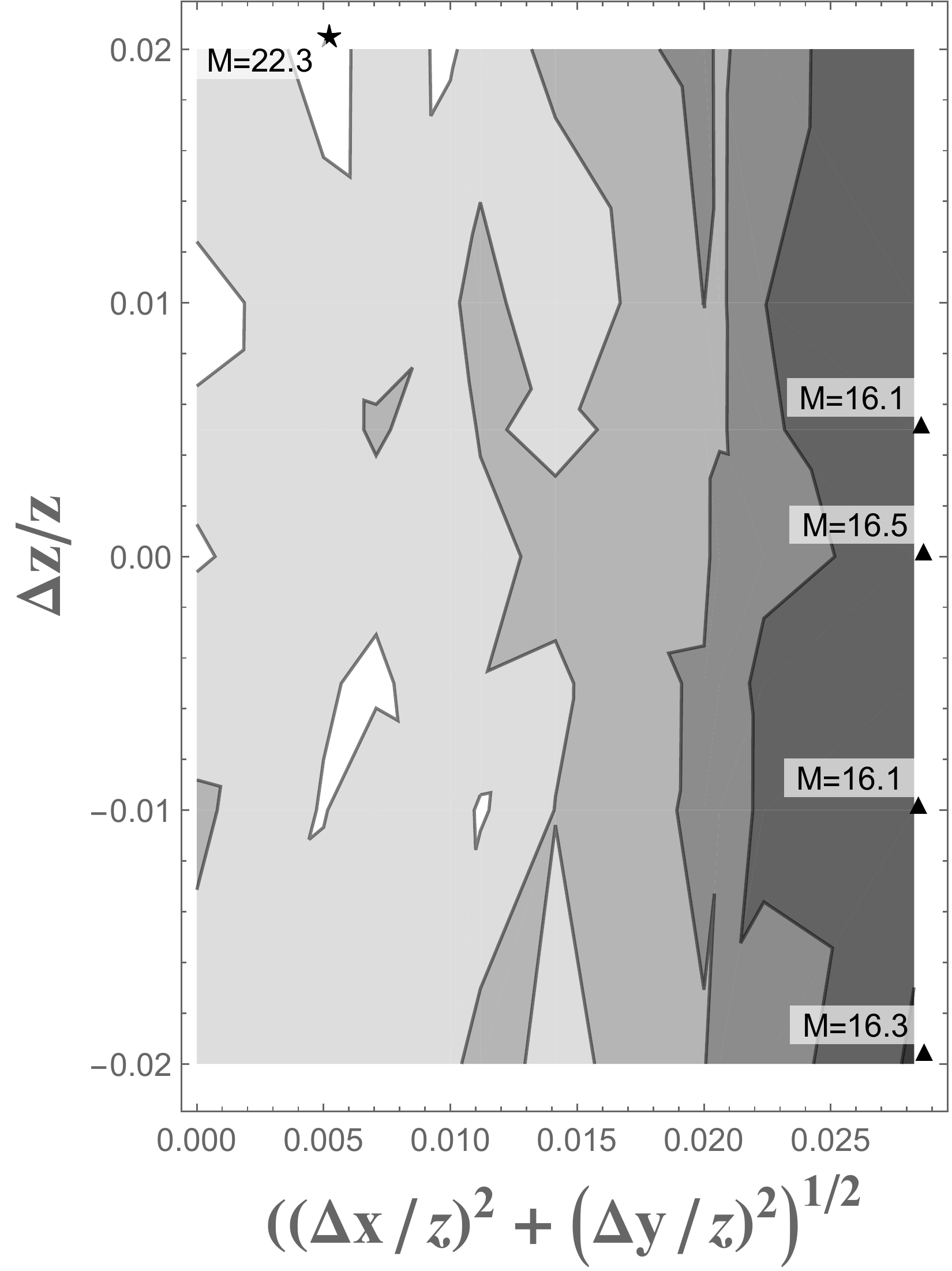}}
\scalebox{0.32}{\includegraphics{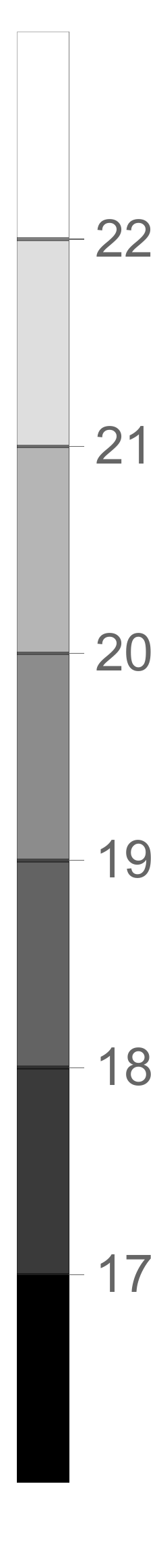}}
\end{tabular}
\end{center}
\caption{
Values of the figures of merit of the optimal indexing solutions in 
(i) cementite case for 19 bands (bandwidths were not used), 
(ii) $Fe$ case for 23 bands (bandwidths were used),
(ii) $Fe$ case for 11 bands (bandwidths were used),
(iv) $Ni$ case for 10 bands (bandwidths were used);
the contours represent the interpolation of the largest $M_{n, N}$ values obtained for the various shifts described in Tables~\ref{Results when band widths are not used}--\ref{Results when band widths are used}.
The $\blacktriangle$ marks indicate the shifts $(\Delta x, \Delta y, \Delta z)$ for which the $M_{n, N}$ values were rather small.
The shifts for which $M_{n, N}$ took the maximum value, are also presented with $\star$ for reference.
The $M_{n, N}$ value can be large even if the initial shift is large, since unit-cell parameters and the projection center are refined after indexing.
}
\label{fig: results for purturbed projection centers}
\end{figure}

Our software stably obtained the correct solutions for any cases in which
$\sqrt{ (\Delta x/z)^2 + (\Delta y/z)^2 + (\Delta z/z)^2 } \le 0.02$.
The results are presented in Tables~\ref{Results when band widths are not used},~\ref{Results when band widths are used}.
The presented unit-cell parameters have been refined by the least-squares method.
If a unit cell close to the correct one obtains the largest $M_{n, N}$ value among the unit cells in the same Bravais type, 
it is counted as a success.
In order to see the influence of imprecise projection centers on the indexing solutions, 
the $M_{n, N}$ values
in (i) the cementite case when bandwidths are not used, and (ii) the $Fe$ case when bandwidths are used,
are presented in Figure~\ref{fig: results for purturbed projection centers}.

When the number of bands used in ab-initio indexing was reduced by half, no correct answer was obtained for the $Zn$ and Cementite samples, even by the exhaustive search. In particular, for the $Zn$ sample, a low-symmetric derivative lattice of the hexagonal lattice was output.
Since all the bands were indexed by the derivative lattice, the $M_{n, N}$ value was still more than 20.

This indicates that the ambiguity problem is more likely to happen, if a smaller number of bands are used.
The $M_{n, N}$ values when the indexing was successful for a smaller number of bands, are presented in (iii), (iv) of Figure \ref{fig: results for purturbed projection centers}. 
As also deduced from the definition, 
the $M_{n, N}$ tends to be larger, when the number of bands to fit is small.

\section{Discussion}

It is known that the de Wolff $M_n$ attains a large value (\EG $> 10$) for very plausible solutions, and does not exceed 3 for invalid solutions \cite{DeWolff68}. 
The following are the other well-known properties of the de Wolff $M_n$:
\begin{enumerate}[(a)]
	\item $M_n$ is sensitive to the existence of reflections observed but not computed from the model, and insensitive to the reflections computed but not observed in the pattern,
because of the asymmetric definition of $M_n$ with regard to the observed and calculated reflection sets. 
	
	\item The value of $M_n$ generally decrease as $n$ increases more than 20, because more unprecise $d$-values are used as a result.

	\item If two unit cells with almost identical parameters but distinct Bravais types are compared, the higher-symmetric cell attains a larger $M_n$, because 
	the peak overlap caused by the symmetry make the number of computed reflections smaller.

\end{enumerate}

According to its definition, $M_{n, N}$ also has the property (a),
which is desirable for use in EBSD indexing, because there are a number of computed but not observed bands in EBSD patterns.
As for (b), $M_{n, N}$ also reflects the accuracy of observation, similarly to the other statistical criteria, such as chi-square values.

Although it is possible to change the definition of $M_{n, N}$ so that it has the property (c), this was not adopted, so
the current $M_{n, N}$ does not possess the property (c), as seen from the values in Table~\ref{Bravais lattice judged from the values of the figure of merit}, because band overlapping occurs regardless of the symmetry.

However, even if the current $M_{n, N}$ is used, plausible solutions with the highest-symmetry can be easily found, just by checking the output list as in Table~\ref{Bravais lattice judged from the values of the figure of merit}.
As seen from Table~\ref{Bravais lattice judged from the values of the figure of merit},
the figures of merit also work well to judge which Bravais type is the true.
For example, in the case of $Ni$, the $M_{n, N}$ values of the cubic (P, I) solutions were much smaller than those provided to the cubic(F) solutions.
The same thing is observed, when the cubic(P, F) and cubic(I) solutions for the $Fe$ pattern, 
and the hexagonal and trigonal solutions for the $Ni$ pattern are compared. 
Considering that all the derivative lattices can index the same band positions and widths,
this is probably owing to our heuristics used when computed reflections are generated, which was described in the last two paragraphs of Section~4.

\onecolumn
\begin{landscape}
\begin{table}[htbp]
\caption{The maximum $M_{n, N}$ values of each Bravais type (attained in the exhaustive searches in Table~\ref{The optimum solutions when band widths are used})
}
\label{Bravais lattice judged from the values of the figure of merit}
\begin{minipage}{\textwidth}
\begin{tabular}{lrrrrrrrrrrp{20mm}}      % Alignment for each cell: l=left, c=center, r=right
\hline
	&	\multicolumn{2}{c}{$\mathbf{Ni}$ (cubic (F))}	& &	\multicolumn{2}{c}{$\mathbf{Fe}$	(cubic (I))}
	& &	\multicolumn{2}{c}{$\mathbf{Zn}$ (hexagonal)} &		& \multicolumn{2}{c}{{\bf Cementite} (orthorhombic (P))} \\
{\bf Bravais type}	&	{\bf Not used}	&	{\bf Used}	& &	{\bf Not used}	&	{\bf Used}	& &	{\bf Not used}	&	{\bf Used}	& &	{\bf Not used}	&	{\bf Used}	\\
\hline
Triclinic	&	51.54$^a$ 	&	23.61 	& &	54.36 	&	34.58 	& &	42.93 	&	23.95 	& &	20.75$^b$ 	&	4.33 	\\
Monoclinic(P)	&	$<$ 36	&	$<$ 12	& &	$<$ 44	&	$<$ 15	& &	39.70 	&	24.00 	& &	20.10$^b$ 	&	4.33 	\\
Monoclinic(C)	&	53.90$^a$ 	&	23.27 	& &	57.35 	&	37.31 	& &	42.49 	&	24.47 	& &	$<$ 10	&	$<$ 3	\\
Orthorhombic(P)	&	$<$ 36	&	$<$ 12	& &	$<$ 44	&	$<$ 15	& &	$<$ 30	&	$<$ 12	& &	16.73$^b$ 	&	4.33 	\\
Orthorhombic(C)	&	$<$ 36	&	$<$ 12	& &	$<$ 44	&	$<$ 15	& &	40.08 	&	24.14 	& &	$<$ 10	&	$<$ 3	\\
Orthorhombic(I)	&	53.01$^a$ 	&	23.26 	& &	52.03 	&	33.99 	& &	$<$ 30	&	$<$ 3	& &	$<$ 10	&	$<$ 3	\\
Orthorhombic(F)	&	50.63$^a$ 	&	22.96 	& &	57.15 	&	36.74 	& &	$<$ 3	&	$<$ 3	& &	$<$ 3	&	$<$ 3	\\
Tetragonal(P)	&	$<$ 36	&	$<$ 12	& &	$<$ 44	&	$<$ 15	& &	$<$ 30	&	$<$ 12	& &	$<$ 3	&	$<$ 3	\\
Tetragonal(I)	&	52.33$^a$ 	&	23.38 	& &	54.55 	&	37.85 	& &	$<$ 30	&	$<$ 3	& &	$<$ 3	&	$<$ 3	\\
Trigonal	&	46.31$^a$ 	&	22.81 	& &	59.14 	&	37.96 	& &	$<$ 30	&	$<$ 12	& &	$<$ 3	&	$<$ 3	\\
Hexagonal	&	$<$ 3	&	$<$ 3	& &	$<$ 44	&	$<$ 15	& &	39.85 	&	24.29 	& &	$<$ 3	&	$<$ 3	\\
Cubic(P)	&	$<$ 36	&	$<$ 12	& &	$<$ 44	&	$<$ 15	& &	$<$ 3	&	$<$ 3	& &	$<$ 3	&	$<$ 3	\\
Cubic(I)	&	$<$ 3	&	$<$ 3	& &	59.03 	&	38.13 	& &	$<$ 3	&	$<$ 3	& &	$<$ 3	&	$<$ 3	\\
Cubic(F)	&	47.56 	&	23.00 	& &	$<$ 44	&	$<$ 3	& &	$<$ 3	&	$<$ 3	& &	$<$ 3	&	$<$ 3	\\
\hline
\end{tabular}
\end{minipage}
	\footnotetext[1]{
Any cubic (F) cell also has the symmetry of tetragonal (F, I), tetragonal and Monoclinic (C).
As seen from this result, relaxed parameters of the correct unit cell frequently attain slightly larger $M_{n, N}$ values than the correct parameters (\EG the triclinic parameters with the largest of $M_{n, N}$ value were
$a:b:c = 1.0 : 1.0 : 1.0$, $\alpha=119.6$, $\beta = 90.5$, $\gamma = 119.9$ (deg.), 
which shows the symmetry of cubic (F)).
Although this can be avoided by slightly changing the definition of the figures of merit,
the correct Bravais type can be estimated even if the current version is used, by comparing the largest $M_{n, N}$ values attained for each Bravais type.
It is rare that some solution with a symmetry higher than the correct one has a large $M_{n, N}$ value.
	}
	\footnotetext[2]{
The parameters with the largest of $M_{n, N}$ values were \\
monoclinic (P): $a:b:c = 0.667 : 0.747 : 1$, $\beta = 91.2$ (deg.), \\
triclinic: $a:b:c = 0.652 : 0.747 : 1$, $\alpha=90.3$, $\beta = 90.3$, $\gamma = 88.9$ (deg.), \\
both of which are close to the correct orthorhombic parameters.
	}
\end{table}
\end{landscape}

%\twocolumn

%It is also easy to verify whether two unit cells of the same centering type are almost identical by comparing their parameters.

\onecolumn
\begin{table}[htbp]
\caption{Indexing result for the cementite pattern; the observed/calculated band positions $(X, Y)$ and bandwidths $\beta$ are compared (the following values have no unit, since the camera lengths is set to 1).}
\label{Indexing result for the Cementite pattern}
\begin{minipage}{\textwidth}
\begin{tabular}{rrrr rrr rrr p{28mm}rr}      % Alignment for each cell: l=left, c=center, r=right
\hline
\multicolumn{4}{r}{{\bf Miller index}} & \multicolumn{2}{r}{$(X^{cal}, Y^{cal})$}     && \multicolumn{2}{r}{$(X^{obs}, Y^{obs})$}     & & 
distance between $(X^{cal}, Y^{cal})$ and $(X^{obs}, Y^{obs})$
& $\beta^{cal}$  & $\beta^{obs}$ \\
\hline
%    1 
&   -2 &   -3 &    3 &   -0.3951 &   -0.0993 &&   -0.3948 &   -0.0985 & %  0.0198 &    0.0085
&    0.0008%     &    1 &         
&0.0937         &0.0892 \\
%    2 
&  $^\dagger$ 0 &    0 &    6 &   -0.0655 &   -0.2868 & &  -0.0649 &   -0.2885 &  %   0.0065 &    0.0186
 &   0.0017 % &    1 &        
 &0.0893         &0.0897 \\
%    3 
&  -3 &   -3 &    0 &   -0.4663 &    0.1836 & &  -0.4653 &    0.1876 &  %   0.0205 &    0.0115
 &   0.0042 % &    1 &        
 &0.1041         &0.0984 \\
%    4 
&   -1 &    0 &   -3 &    0.0090 &    0.1526 & &   0.0088 &    0.1499 &   %  0.0028 &    0.0178
 &   0.0027 % &    1 &         
&0.0470         &0.0401 \\
%    5 
&   -2 &   -3 &   -3 &   -0.0600 &    0.0654 &&   -0.0615 &    0.0665 &  %   0.0120 &    0.0130
 &   0.0019 % &    1 &         
&0.0808         &0.0823 \\
%    6 
&   -1 &    0 &    3 &   -0.3670 &   -0.7893 & &  -0.3626 &   -0.7930 &  %   0.0188 &    0.0286
 &   0.0058 % &    1 &         
&0.0813         &0.0757 \\
%    7 
&    0 &    4 &   -3 &    0.1443 &    0.0584 &  &  0.1477 &    0.0591 &    % 0.0166 &    0.0071
 &   0.0034 % &    1 &         
&0.0861         &0.0800 \\
%    8 
&    1 &    4 &    0 &    0.0775 &   -0.0141 &  &  0.0790 &   -0.0154 &   %  0.0172 &    0.0036
 &   0.0021 % &    1 &        
 &0.0767         &0.0739 \\
%    9 
&   -1 &    4 &    0 &    0.6853 &   -0.0091 & &   0.6879 &   -0.0069 &   %  0.0257 &    0.0120
 &   0.0034 % &    1 &         
&0.1126         &0.1118 \\
%   10 
&   -1 &    2 &    5 &    0.1794 &   -0.7803 & &   0.1845 &   -0.7818 &   %  0.0152 &    0.0281
 &   0.0053 % &    1 &         
&0.1328         &0.1331 \\
%   11 
&    1 &    2 &    5 &    0.0536 &   -0.1553 &  &  0.0532 &   -0.1549 &   %  0.0064 &    0.0170
 &   0.0006 % &    1 &         
&0.0826         &0.0816 \\
%   12
 &    0 &    4 &    3 &    0.3773 &   -0.2902 & &   0.3747 &   -0.2907 &  %   0.0176 &    0.0146
 &    0.0026 % &    1 &        
  &0.1035         &0.0990 \\
%   13 
&   -1 &   -2 &    5 &   -0.2421 &   -0.2327 &&   -0.2415 &   -0.2352 & %    0.0145 &    0.0142
 &   0.0026 % &    1 &         
&0.0896         &0.1095 \\
%   14 
&   -1 &   -2 &    1 &   -0.2346 &   -0.0082 &&   -0.2334 &   -0.0072 & %    0.0184 &    0.0041
 &   0.0016 % &    1 &         
&0.0467         &0.0526 \\
%   15 
&  $^\ddagger$0 &    2 &   -2 &    0.0870 &    0.0486 &  &  0.0894 &    0.0482 &    % 0.0155 &    0.0085
 &   0.0025 % &    1 &         
&0.0462         &0.0440 \\
%   16 
&   -1 &    2 &   -5 &    0.0770 &    0.1222 & &   0.0758 &    0.1206 &   %  0.0097 &    0.0151
 &   0.0020 % &    1 &         
&0.0820         &0.0872 \\
%   17 
&    0 &    2 &    5 &    0.1253 &   -0.4276 &  &  0.1219 &   -0.4292 &   %  0.0094 &    0.0202
 &   0.0038 % &    1 &         
&0.0933         &0.0874 \\
%   18 
&   -1 &   -1 &    3 &   -0.4078 &   -0.3758 &&   -0.4045 &   -0.3765 & %    0.0179 &    0.0171
 &   0.0033 % &    1 &         
&0.0649         &0.0741 \\
%   19
 &   $^\ddagger$0 &    2 &    2 &    0.3381 &   -0.3489 & &   0.3388 &   -0.3445 &  %   0.0162 &    0.0164
 &    0.0045 %  &    1 &         
 &0.0567         &0.0617 \\
\end{tabular}
	\footnotetext[2]{
From the reflection rules of $P\ b\ n\ m$ (No.62),
$\{ 0 0 \ell \}$ ($\ell$: odd) may be excluded.
	}
	\footnotetext[3]{
$\{ 0 1 \bar{1} \}$ and $\{ 0 1 1 \}$ were forbidden by 
the reflection rules ($0 k l$ with an odd $k$) of $P\ b\ n\ m$ (No.62).
	}
\end{minipage}
\end{table}

%\twocolumn

Table~\ref{Indexing result for the Cementite pattern}
is the indexing result for the Cementite sample.
The bandwidths assigned the Miller indices 
$( 0 2 \bar{2} )$ and $( 0 2 2 )$ were probably due to SA, considering that $( 0 1 \bar{1} )$ and $( 0 1 1 )$ are forbidden by the rules ($0 k l$ with an odd $k$) of $P\ b\ n\ m$ (No.62).
However, influence of non-visible narrowest band widths (and also underestimation of the unit-cell scale) is also observed from the 
assigned Miller indices (\EG $( 0 0 6 )$ and $( \bar{3} \bar{3} 0 )$).
%In the simulated patterns, all the input bandwidths were the narrowest ones.

With regard to the ambiguity pointed out for the first time in this article,
when bandwidths are used, it is mainly caused by sublattices $M$ of the true crystal lattice $L$ with a small index $[L : M]$,
because $M$ can index all the edges of the narrowest bands of $L$.
However, if the HOLZ rings are used, 
ambiguity is mainly caused by superlattices $M$ of $L$ with small $[M : L]$,
because the radius of a HOLZ ring is an integer multiple of 
the shortest length of lattice vectors perpendicular to the corresponding zone \cite{Michael2000b}.
This indicates that 
the entire information contained in an EBSD pattern might be able to resolve the ambiguity.

\section{Conclusion}

For ab-initio indexing, a new method based on distribution rules of systematic absence and error-stable Bravais lattice determination
was proposed. 
In addition, the de Wolff figures of merit for 1D powder diffraction patterns 
were redefined for data of multiple dimension and used in orientation determination and ab-initio indexing.
%The indexing result for an experimental cementite pattern indicates that 
%indexing can be successful, even if precise projection centers are not given (Table~\ref{}).
The new figures of merit have properties similar to those of the original de Wolff $M_n$,
except for the preference for higher-symmetric cells, although it is possible to change the definition of $M_n$ so that it has such a preference.
Even if the current figures of merit are used, users can efficiently find the optimal solution and Bravais type.
It was also explained how erroneous band widths can cause ambiguity of solutions, in particular in case of low-symmetric cells.

%For future studies, 
%combinatorial use of band widths and the HOLZ rings was proposed, in order to have more reliable grounds in 
%analytical results.

     %-------------------------------------------------------------------------
     % The back matter of the paper - acknowledgements and references
     %-------------------------------------------------------------------------

     % Acknowledgements come after the appendices

\paragraph{Acknowledgments}
This study was financially supported by the PREST (JPMJPR14E6). 
We would like to extend our gratitude 
to Dr.~A.~Esmaeili, Ms.~T.~Ueno of Yamagata University
and Mr.~S.~E.~Graiff-Zurita of Kyushu University, who helped us in coding the software, preparing the input files, and performing the computation.

     % References are at the end of the document, between \begin{references}
     % and \end{references} tags. Each reference is in a \reference entry.

%\begin{references}
%\reference{Author, A. \& Author, B. (1984). \emph{Journal} \textbf{Vol}, 
%first page--last page.}
%\end{references}

     %-------------------------------------------------------------------------
     % TABLES AND FIGURES SHOULD BE INSERTED AFTER THE MAIN BODY OF THE TEXT
     %-------------------------------------------------------------------------

     % Simple tables should use the tabular environment according to this
     % model

\bibliographystyle{apalike}
\bibliography{iucr}

     % Postscript figures can be included with multiple figure blocks

\end{document}